%% file: jness.tex
\documentclass{aa}
\usepackage{graphics}

\usepackage{natbib}
\usepackage{graphicx}

\begin{document}

\title{Are stellar coronae optically thin in X-rays?}
\subtitle{A systematic investigation of opacity effects}

\author{J.-U. Ness\inst{1}, J.H.M.M. Schmitt\inst{1}, M. Audard\inst{2}, M. G\"udel\inst{3},
R. Mewe\inst{4}}
\institute{
Hamburger Sternwarte, Universit\" at Hamburg, Gojenbergsweg 112,
D-21029 Hamburg, Germany
\and
Columbia Astrophysics Laboratory, 550 West 120th Street, New York, NY 10027, USA
 \and
Paul Scherrer Institut, W\"urenlingen \& Villigen, 5232 Villigen PSI, Switzerland
 \and
National Institute for Space Research (SRON), Sorbonnelaan 2, NL-3584 CA
Utrecht, the Netherlands}

\authorrunning{Ness, Schmitt, Audard, et al.}
\titlerunning{Are stellar coronae optically thin?}
\offprints{J.-U. Ness}
\mail{jness@hs.uni-hamburg.de}
\date{Received January 23, 2003; accepted June 06, 2003}

\abstract{
The relevance of resonant scattering in the solar corona has always been
discussed controversially. Ratios of emission lines from identical ions but
different oscillator strengths have been used in order to estimate damping of
resonance lines due to possible resonant scattering, i.e., absorption by
photo-excitation and re-emission out of the line of sight. The analysis of
stellar spectra in analogy to previous works for the Sun is possible now with
XMM-Newton and Chandra grating spectra and requires this issue
to be considered again. In this work we present a sample of 44 X-ray spectra
obtained for 26 stellar coronae with the RGS on board XMM-Newton and the LETGS
and HETGS on board Chandra. We use ratios of the Fe\,{\sc xvii} lines at
15.27\,\AA\ and 16.78\,\AA\ lines to the resonance line at 15.03\,\AA\ as well
as the He-like f/r ratio of O\,{\sc vii} and Ne\,{\sc ix} to measure optical
depth effects and compare them with ratios obtained from optically thin plasma
atomic databases such as MEKAL, Chianti, and APEC. From the Fe\,{\sc xvii} line
ratios we find no convincing proof for resonance line scattering. Optical
depths are basically identical for all kinds of stellar coronae and we conclude
that identical optical depths are more probable when effects from resonant
scattering are generally negligible. The 15.27/15.03\,\AA\ ratio shows a
regular trend suggesting blending of the 15.27\,\AA\ line by a cooler Fe line,
possibly Fe\,{\sc xvi}. The He-like f/r ratios for O and Ne show no indication
for significant damping of the resonance lines.\\
We mainly attribute deviations from the atomic databases to still uncertain
emissivities which do not agree well with laboratory measurements and which come
out with differing results when accounting for one or the other side effect.
We attribute the discrepancies in the solar data to geometrical effects from
observing individual emitting regions in the solar corona but only overall
emission for stellar coronae including photons eventually scattered into the
line of sight.
\keywords{stars: activity -- stars: coronae -- stars: late type -- X-rays: stars -- atomic processes: resonant scattering}
}
\maketitle

\section{Introduction}
\label{intro}

The emission line spectra obtained with the gratings on board the new X-ray
observatories XMM-Newton and Chandra allow us to measure individual X-ray emission
lines originating from ions in high ionization stages. These emission lines
probe the hot tenuous plasma in stellar coronae. Obviously, the solar corona
is much easier to study than stellar coronae, and observing techniques and
methods originally developed for the analysis of the solar corona can now be
applied to stellar coronae many years later with much improved technology.
The theory required for X-ray spectroscopy developed in the 1960s and 70s now
experiences a revival with the new generation of X-ray instruments applied to
study stellar coronae.

\input{tab1}

The basic assumptions underlying almost all theoretical and observational
analyses of solar and stellar coronal emission lines are, first, that the plasma
is optically thin, and second, that the plasma is in collisional equilibrium.
The latter implies that excitations are exclusively due to collisions and not to
photo-excitation, the former implies that all photons produced in the hot plasma
escape without further interaction. The plasma then cools through radiation
(and possibly conduction). Radiative transport does not need to be considered,
which makes the interpretation of coronal spectra and modeling of the coronal
plasma much easier.\\
If this assumption were not true, opacity effects would first become visible in
strong resonance lines. Resonance line photons could be absorbed and
re-emitted in other directions. Depending on the plasma geometry, resonance
line photons can be scattered out of the line of sight, but photons can also be
scattered from other directions into the line of sight. Photons scattered back
to the stellar surface will be absorbed rather than escape. Therefore, the line
intensities of lines with strong scattering are reduced when compared to lines
with no scattering.
This effect is called resonant scattering. In coronal equilibrium forbidden
lines can always be considered optically thin because of their low radiative
transition probabilities. Therefore the effect of resonant scattering can be
recognized by resonance lines being damped in comparison to forbidden lines.
Thus, the basic principle for detecting resonance scattering is to measure line
flux ratios of definitely non-damped forbidden lines with low oscillator
strengths $f$ and resonance lines with high oscillator strengths. If this
ratio is found to be enhanced compared to line ratios from theoretical
predictions or from laboratory measurements, the resonance line should be
considered optically thick. For a detailed account of the underlying theory
we refer to \cite{bhatia01,schmelz97,mar92}. We derive the reference
line ratios from the line databases MEKAL\footnote{Improved version; available
at http://www.sron.nl/\\
divisions/hea/spex/version1.10/line/line\_new.ps.gz} \citep{mewe95}, Chianti
with ionization balances from \cite{ar} \citep{dere01,young02}, and
APEC\footnote{Version 1.2; available at http://cxc.harvard.edu/atomdb}
\citep[e.g.,][]{smith01}.\\
In the solar context the problem of resonant scattering of X-ray emission lines
has been discussed with rather controversial conclusions. \cite{actcat76},
\cite{acton78}, and \cite{strong78} investigated the effects
of resonant scattering for various He-like ions, especially the
O\,{\sc vii} resonance line at 21.6\,\AA. They found differences between
theoretical and observed values of the temperature sensitive G-ratio (f+i)/r
\citep{gj69} and interpreted these differences as being due to resonant
scattering effects. \cite{schmelz97} and \cite{saba99} measured five different
line ratios and found significant optical depths only for the Fe\,{\sc xvii}
line at 15.03\,\AA\ ($^1$S$_0\rightarrow^1$P$_1$ with a high oscillator
strength $f=2.66$). They compared the 15.03\,\AA\ line flux with
line flux measurements for Fe\,{\sc xvii} lines with lower oscillator strengths,
namely two intercombination lines $^1$S$_0\rightarrow^3$D$_1$ at 15.27\,\AA\ 
with $f=0.593$ and $^1$S$_0\rightarrow^3$P$_1$ at 16.78\,\AA\ with $f=0.1$.
The different oscillator strengths indicate to which extent the transition can
be subject to resonant scattering, i.e., the probability for resonant scattering
of the 15.27\,\AA\ line is less than a quarter of that of the 15.03\,\AA\ line,
while resonant scattering of the 16.78\,\AA\ line is even less probable, i.e.,
0.04 times that for the 15.03\,\AA\ line. For the prediction of such line
ratios for optically thin cases theory
and experiment unfortunately do not agree with each other. The 15.27/15.03\,\AA\
line ratio has been measured in the Electron Beam Ion Trap
\citep[EBIT;][]{brown98,brown01,laming00}. These experiments typically yield
Fe\,{\sc xvii} 15.27/15.03\,\AA\ photon flux ratios in the range 0.3 - 0.36,
which significantly differ from those expected from theoretical calculations.
Also, \cite{brown01,phil97} point out that contamination of the 15.27\,\AA\
line by an Fe\,{\sc xvi} satellite line can further enhance the observed
15.27/15.03\,\AA\ photon flux ratio especially in cooler plasmas (below $\sim
3$\,MK).\\
The optical thickness of stellar coronae has been investigated for EUV lines
\citep[e.g.,][]{schr94,schm96} measured with the Extreme Ultraviolet Explorer
(EUVE). While \cite{schr94} claim to have found evidence for resonant
scattering, \cite{schm96} argue using additional ROSAT observations that
resonant scattering does not appear to be required for the interpretation of
the EUV and X-ray spectra of inactive cool stars. From Chandra LETGS
measurements \cite{ness01} ruled out optical depth effects in their
analysis of Procyon and Capella. The assumption of a significant optical depth
leads to unreasonably large emission measures contradicting their direct
measurements of emission measures. \cite{mewe01} measured the Fe\,{\sc xvii}
15.27/15.03\,\AA\ photon flux ratio for Capella of $0.35\,\pm\,0.02$ and derive
a formal value of an optical depth $\tau$ (assuming slab geometries), which can
be used in order to constrain loop lengths. The effects of opacity effects
have also been addressed for Capella by \cite{phil01} using the same ratios
and were found to be neglible. \cite{ness_alg} measure the same
ratio identical to the Capella measurement for the much more active star Algol.
From this consistency they conclude that resonant scattering effects might in
general be negligible for all coronae rather than being identical for all kinds
of different coronae. This hypothesis is
also supported by \cite{aud03} from an analysis of a sample of five active
RS CVn stars, where also similar ratios are measured for all stars.\\
The purpose of this paper is a systematic investigation of potential optical
depth effects in a large sample of stars covering a wide range of different
activity levels. We will specifically analyze two Fe\,{\sc xvii} line ratios
and He-like f/r ratios for O\,{\sc vii} and Ne\,{\sc ix} for all cool stars,
for which high-resolution spectra with the new X-ray instruments
are available. We analyze 22 spectra obtained with the Reflection Grating
Spectrometer (RGS) on board XMM-Newton, 12 spectra measured with the Low Energy
Transmission Grating (LETGS) on board Chandra, and 10 spectra from the High
Energy Transmission Grating (HETGS) on board Chandra (which are split in two
spectra, the Medium Energy Grating (MEG) with a higher aperture and the High
Energy Grating (HEG) with higher spectral resolution). Some stars have been
measured by two or three instruments allowing comparison of calibration and/or
finding variability of opacity effects. We will discuss possible trends and
agreement and disagreement for measured line ratios with theoretical
predictions. The major question we address is: Are resonant scattering effects
dependent on the degree of activity, or are they negligible?

\section{Reduction and analysis}

\subsection{Reduction of the raw data}

For a most comprehensive analysis we studied line ratios relevant for detecting
opacity effects from different instruments. From the XMM-Newton RGS GT program
on board XMM-Newton, 22 spectra from stars in all stages of coronal
activity are available. The reduction procedure for these data is identical
for all spectra using SAS version 5.2. Five stars in our sample (RS~CVn systems)
have been described by \cite{aud03} and a detailed description of the reduction
is given there. For some stars (47 Cas, AU Mic, $\kappa$ Cet, and YZ CMi) we
tested the effect of larger extraction regions comprising 95\% source photons
(instead of 90\%), but find no significant improvement. Three observations
(AB Dor, Capella, and EQ Peg) have been carried out before the chip failure on
the RGS1, so that the range between 10.5 and 13.8\,\AA\ is available also
with the RGS1 for these stars. The analysis of Ne\,{\sc ix} is still not
possible with the RGS1 for these stars, because of bad pixels on the chip
where the photons from the 13.7\,\AA\ (the Ne\,{\sc ix} forbidden line) are
extracted.
Line counts are measured with the CORA program (Sect.~\ref{lfxlues}) and
the ASCII files required for CORA were produced with XSPEC from the fits files
returned by the SAS software. From the response matrices effective areas were
calculated and stored in ASCII files which are used as look-up tables for
converting measured line counts into line fluxes.\\ 
Most of the LETGS data included have been introduced by \cite{ness_10} and for
details on the data reduction we refer to that paper (effective areas from Deron
Pease, Aug. 2002). We also analyze HETGS spectra of all cool stars available
to us and use the pre-processed pha files from the Chandra archive. In
Table~\ref{sample} we list specifications for 44 observations of 26 stars with
exposure times and
X-ray luminosities obtained from the different instruments. We summed all first
order photons converted to energy fluxes using the effective areas, exposure
times and distances in order to calculate X-ray luminosities. Differences in
X-ray luminosities by no more than a
factor of two occur, although they are extracted in the same wavelength
intervals (except for MEG and HEG, which are extracted in their complete
wavelength ranges), because photons in the chip gaps on RGS1 and RGS2 are
missing and higher order photons in the LETGS are not corrected for.\\

\subsection{Measurement of line fluxes}
\label{lfxlues}

Line counts are measured with a modified version of the CORA program by
\cite{newi}. Due to small count numbers all spectra in our sample require
Poisson statistics to be applied. Since the conventional background
subtraction ruins the Poissonian statistics, we construct a model spectrum
consisting of the sum of three components. The line spectrum is modeled with
analytical line profile functions representing instrumental point spread
functions (Lorentzian for the RGS spectra, Gaussian for the MEG and HEG
spectra and a ''$\beta$ model'' for the LETGS spectra, which is a Lorentzian
with an exponent $\beta=2.5$). The background is split in two components, the
instrumental background (extracted from regions on the detectors adjacent to
the dispersion directions) and a source continuum (modeled as a constant value
representing a number of counts per bin over the wavelength region under
individual consideration). The sum of these three components is compared to
the non-subtracted spectrum in order to calculate likelihood values to be
minimized. The modeling is restricted only to the line parameters' position,
line width and line counts, but the two background components must be given
a priori (cf. Sect.~\ref{bg}). Therefore the errors (1\,$\sigma$ errors) given
for the line counts represent only statistical errors (including correlated
errors from possible line blends), but systematic uncertainties from the
placement of a continuum value are not included.

\subsection{Placement of the continuum}
\label{bg}

The accuracy of the iterated line model clearly depends on the choice of the two
background components. The instrumental background is no problem, because it
can be measured from adjacent regions on the detector plates. However, the
determination of reliable source continua (comprising true continuum and
pseudo continuum of unresolved weak lines) is much more difficult. We consider
the source continuum to be constant over small wavelength regions (small range
including the emission
lines to be measured) and assign a single source background parameter $sbg$ in
units counts/\AA\ to represent this flat source continuum. In the CORA program
such a value for a source continuum can be specified directly by hand or the
median value of all bins in the wavelength region covering the lines under
consideration can be selected, which is only valid as long as less than 50\% of
the bins belong to emission lines. All bins containing count numbers higher than
3\,$\sigma$ above this median value ($\sigma=\sqrt{\rm median}$) are regarded
to obviously belong to emission lines and are excluded from calculating the
final $sbg$ median value.\\
The specific challenge posed by RGS spectra is that the line wings are
broad and overlap. The inclusion of correlated statistical errors is thus very
important, but the determination of an adequate value for the source background
is more difficult. The median value will systematically overestimate the
source continuum, because more bins belong to the emission lines rather than
representing the source continuum and line counts will then be underestimated.
For the purpose of this paper, the Fe\,{\sc xvii} lines around 15\,\AA\ are
measured, and this wavelength region contains many nearby lines, such that
for the median calculation only small regions representing the continuum are
available.\\

We therefore modified the program to calculate a value for the source
background by refining the median calculation. The 3\,$\sigma$ criterion
is already an attempt to remove some bins that belong to emission lines in
order to increase the percentage of remaining bins belonging to the background.
For our purpose we modify this criterion in two ways. First, the removal of
bins with high count numbers and the calculation of a new median value are
repeated iteratively until no more bins contain more counts than 3\,$\sigma$
above the respective median values. Secondly, a new parameter $n_\sigma$ is
introduced. In this way median values are iteratively calculated after removal
of all bins with count values higher than $n_\sigma\times \sigma$, i.e.,
median$_{\rm new}=$median(bins$ < n_\sigma\times \sqrt{{\rm median}_{\rm last}})$.
Small values of $n_\sigma$ will more critically remove high-count bins
resulting in lower source background values. Usage of this parameter represents
a parameterized choice of source continuum values by eye.

\begin{figure}[ht]
  \resizebox{\hsize}{!}{\includegraphics{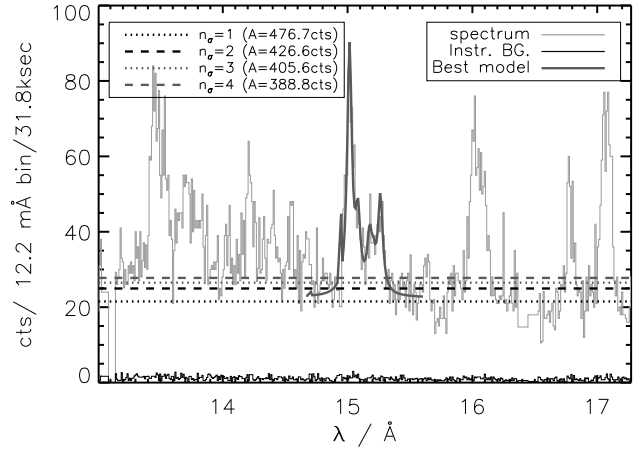}}
\caption{\label{bg15}15\,\AA\ region of $\lambda$ And with RGS2. Placement of a
constant source continuum with median values using different parameters
$n_\sigma$. The respective counts resulting from the different choices of
$n_\sigma$ are given in the upper left. The best model with four lines is
obtained for $n_\sigma=1$ and is overplotted with dark grey. An overestimated
continuum value leads to systematically underestimated 15.27/15.03\,\AA\ line
count ratios hiding possible resonant scattering effects.}
\end{figure}

In Fig.~\ref{bg15} we show the 15\,\AA\ region of $\lambda$~And with attempts
to obtain a most realistic source background value using the new parameter.
It can be seen that this wavelength region is full of emission lines and that
significantly more than 50\% of all bins belong to emission lines rather than
the background emission. By gradually reducing $n_\sigma$ the median background
can significantly be reduced, and with $n_\sigma=1$ a most suitable background
is found. The resulting count number for the 15.03\,\AA\ line range from 388.8
to 476.7 counts. This demonstrates that systematic errors of order 25\% must
be added to the given statistical errors. In the following we use $n_\sigma=1$
for all spectra when fitting the 15\,\AA\ lines and $n_\sigma=1.5$ for the
16.78\,\AA\ line. The neon and oxygen lines are all measured with the old method.

\subsection{Measured line counts}

Since in the RGS1 bad pixels corrupt the measurement of the 15.27\,\AA\ line
we analyze only the RGS2 data and the LETGS, MEG, and HEG data for the iron
measurements. The He-like lines were measured with RGS1, LETGS, and MEG (oxygen)
and with RGS2, LETGS, MEG, and HEG (neon).

\begin{figure}[t]
  \resizebox{\hsize}{!}{\includegraphics{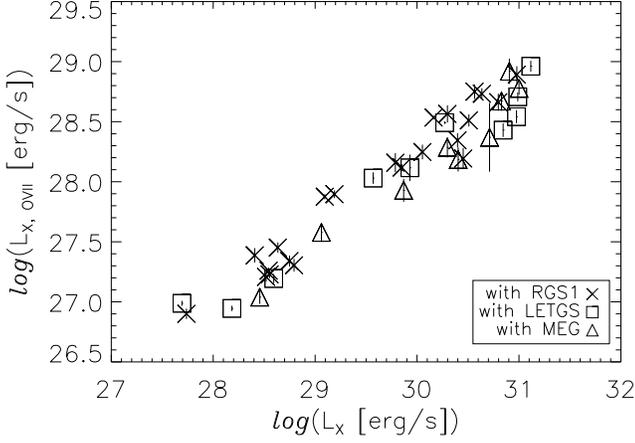}}
\caption{\label{lxox}O\,{\sc vii} line fluxes converted to luminosities
compared to total X-ray luminosities: The O\,{\sc vii}-luminosities are
representative as activity indicators.}
\end{figure}

\input{tab2}

The fit results for the three Fe\,{\sc xvii} lines at 15.03\,\AA, 15.27\,\AA,
and at 16.78\,\AA\ are listed in Table~\ref{fe_cts}.
These counts are converted to energy fluxes in order to derive line
flux ratios using effective areas obtained from the response matrices for
comparison with line flux ratios from the databases MEKAL, Chianti, and APEC,
which all list optically thin emissivities for given temperature grids. The
results for the measured ratios are also listed in Table~\ref{fe_cts}.

\input{tab3}

The line counts measured for the He-like f and r lines of O\,{\sc vii} and
Ne\,{\sc ix} are listed in Tables~\ref{fr_ox} and \ref{fr_ne}, respectively.
We also measure the O\,{\sc viii} Ly$_{\alpha}$ line, and from the O\,{\sc viii}
Ly$_\alpha$/O\,{\sc vii}\,r line ratios we assign an characteristic coronal
temperature to each star (using the APEC line database). Further, we calculate
X-ray luminosities emitted in all three He-like lines (r, i, and f summed) as
activity indicators (cf. Fig.~\ref{lxox}). In addition to the line counts for
oxygen we list the derived temperatures and O\,{\sc vii} and Ne\,{\sc ix}
luminosities in Tables~\ref{fr_ox} and \ref{fr_ne}. The plasma temperature is
also a good activity indicator \citep{gdl97}.

\input{tab4}

\section{Results}
\label{results}

The measured line fluxes are used in order to plot the line ratios sensitive
to resonant scattering versus activity indicators, i.e., temperatures for the
Fe\,{\sc xvii} ratios and X-ray luminosities contained in the He-like lines
for the f/r ratios of O\,{\sc vii} and Ne\,{\sc ix}.\\

\begin{figure*}[!ht]
  \resizebox{\hsize}{!}{\includegraphics{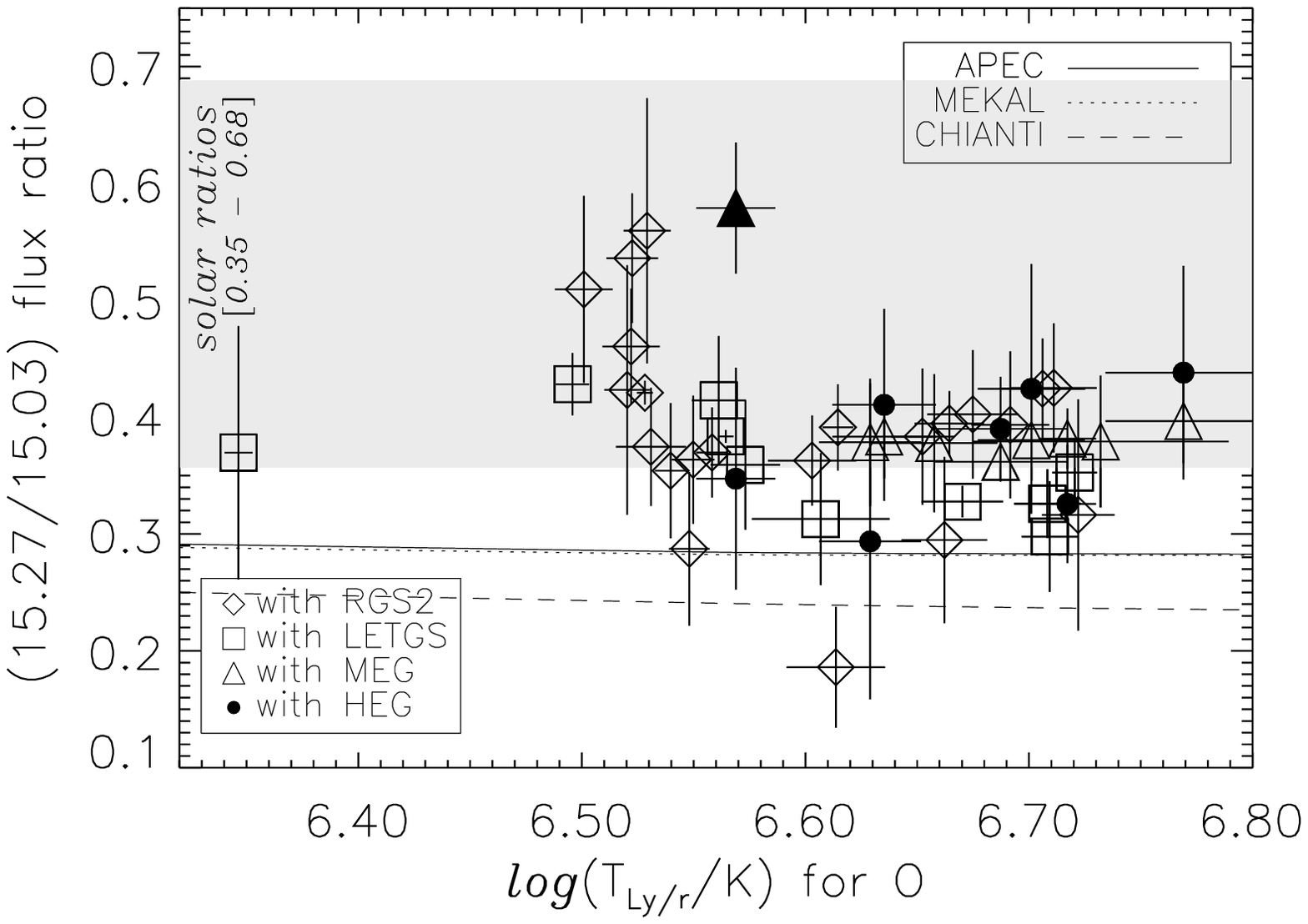}\includegraphics{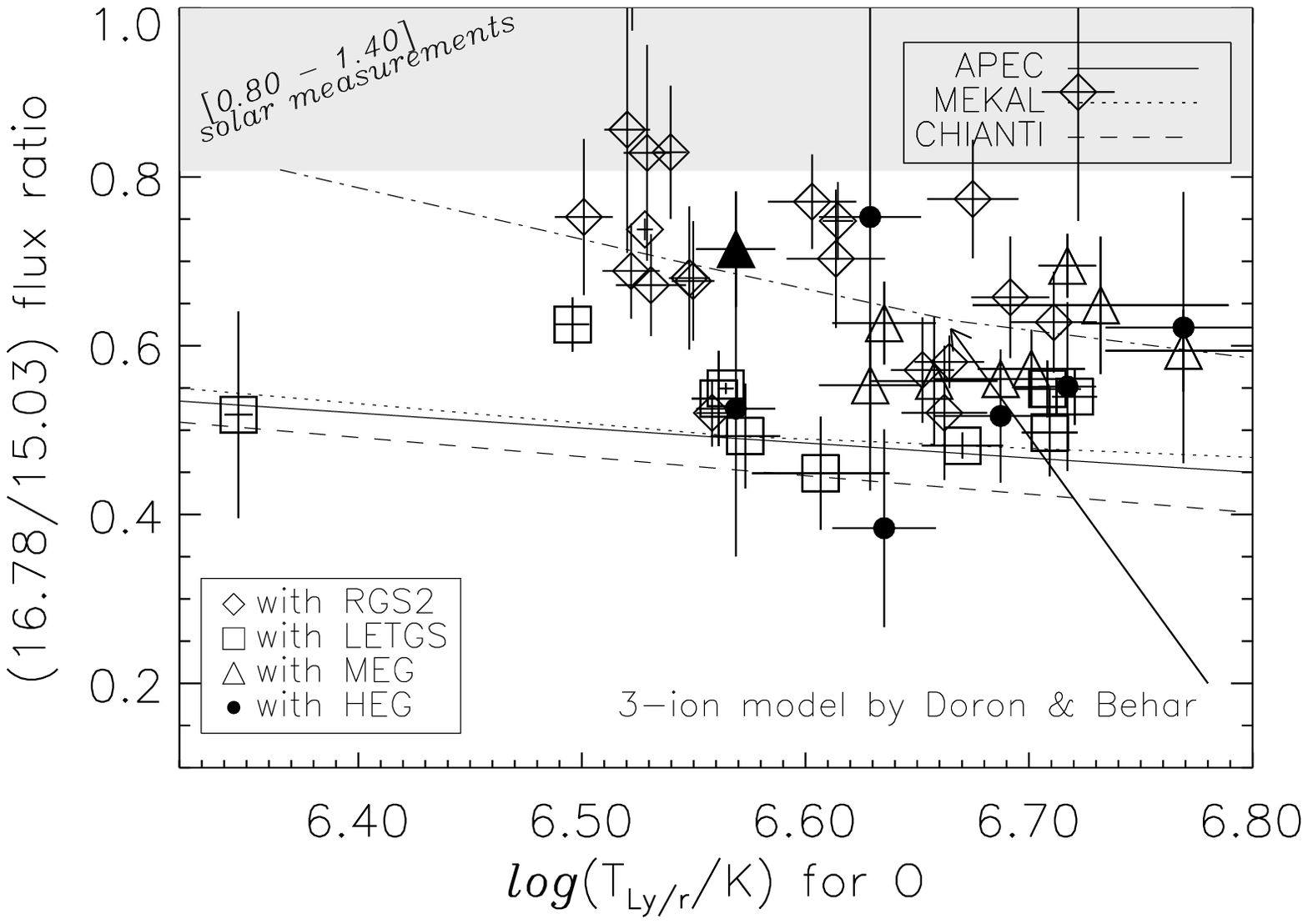}}
\caption{\label{feratios}Investigation of opacity effects for the Fe\,{\sc xvii}
resonance line ($\lambda=15.03$\,\AA, $f=2.66$) as a function of characteristic
temperatures derived from O\,{\sc viii}/O\,{\sc vii} (Ly$_\alpha$/r)
line ratios. {\bf Left panel:} Line ratios with the Fe\,{\sc xvii} line at
15.27\,\AA\ ($f=0.593$), theoretical (low-optical depth) ratios from
APEC, Chianti, and MEKAL, and solar measurements from \cite{saba99} (shaded
area). {\bf Right panel:} Line ratios with the Fe\,{\sc xvii} line at
16.78\,\AA\ ($f=0.01$). Same as in the left panel with additionally new
calculations by \cite{doron}.}
\end{figure*}

\begin{figure*}[!ht]
  \resizebox{\hsize}{!}{\includegraphics{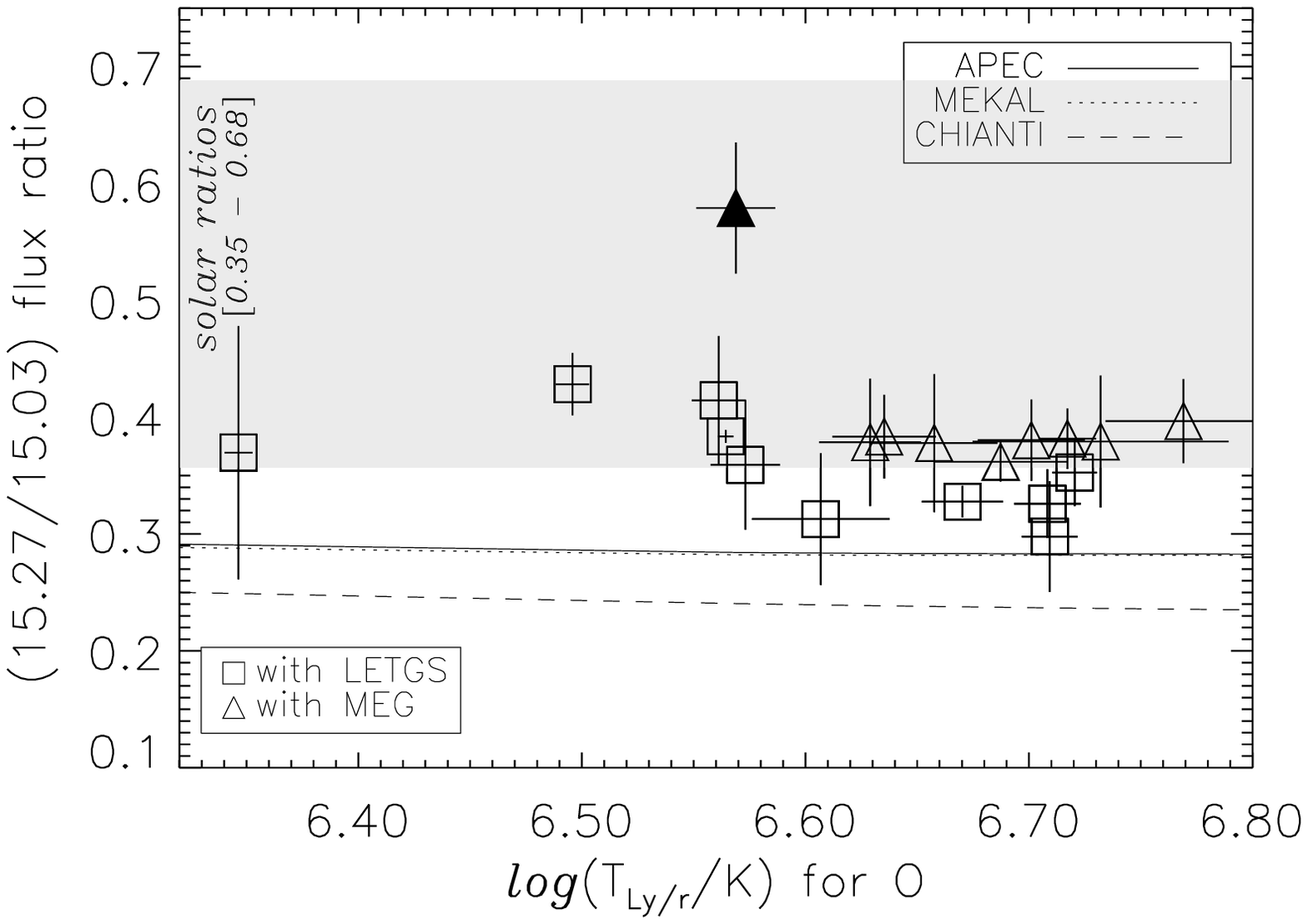}\includegraphics{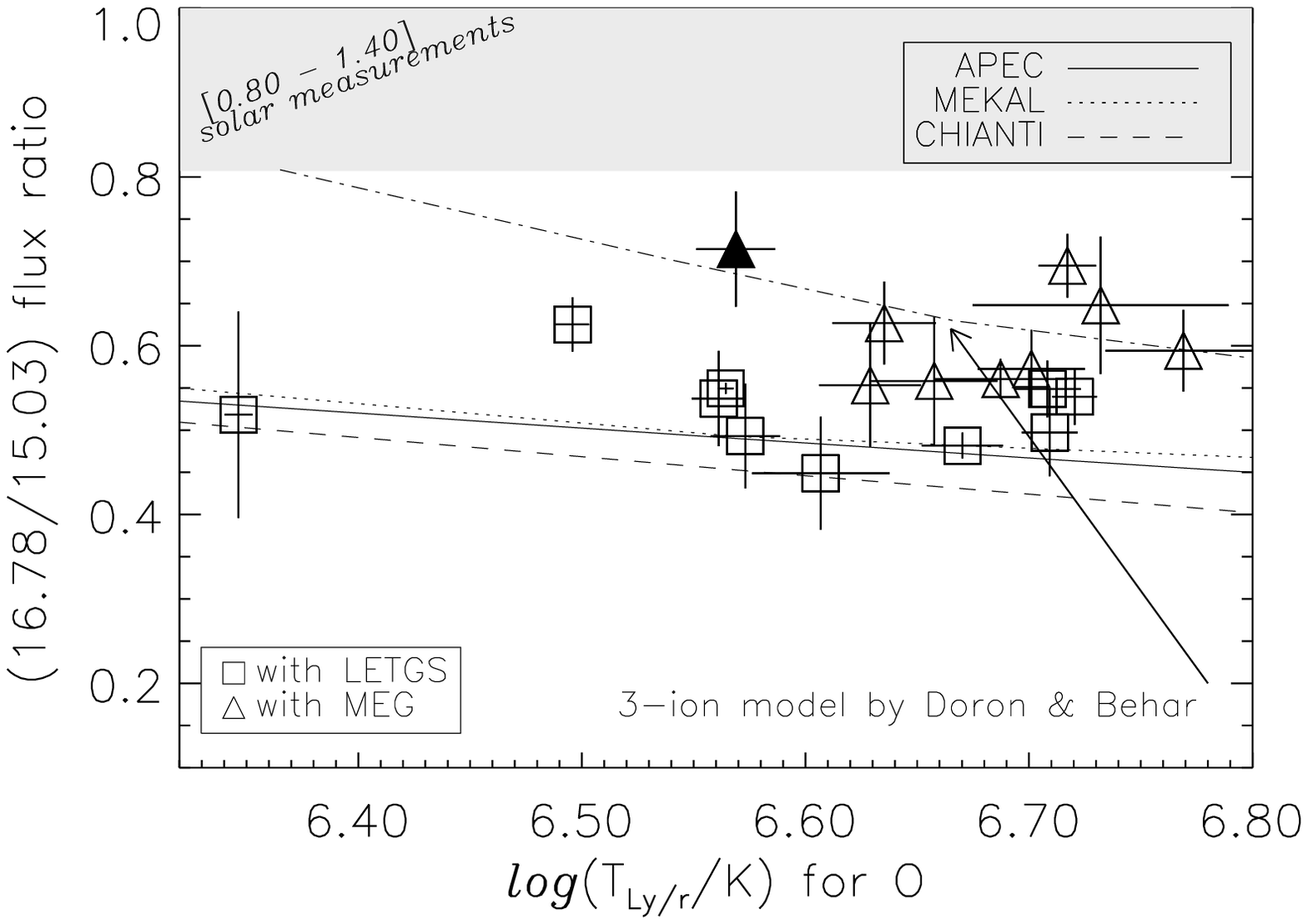}}
\caption{\label{fe_meg}Same as Fig.~\ref{feratios} with only the MEG and
LETGS measurements. The only ratio deviating from the others is measured for
EV Lac (see spectrum in Fig.~\ref{evlac}) marked with a filled triangle.}
\end{figure*}

\begin{figure*}[!ht]
  \resizebox{\hsize}{!}{\includegraphics{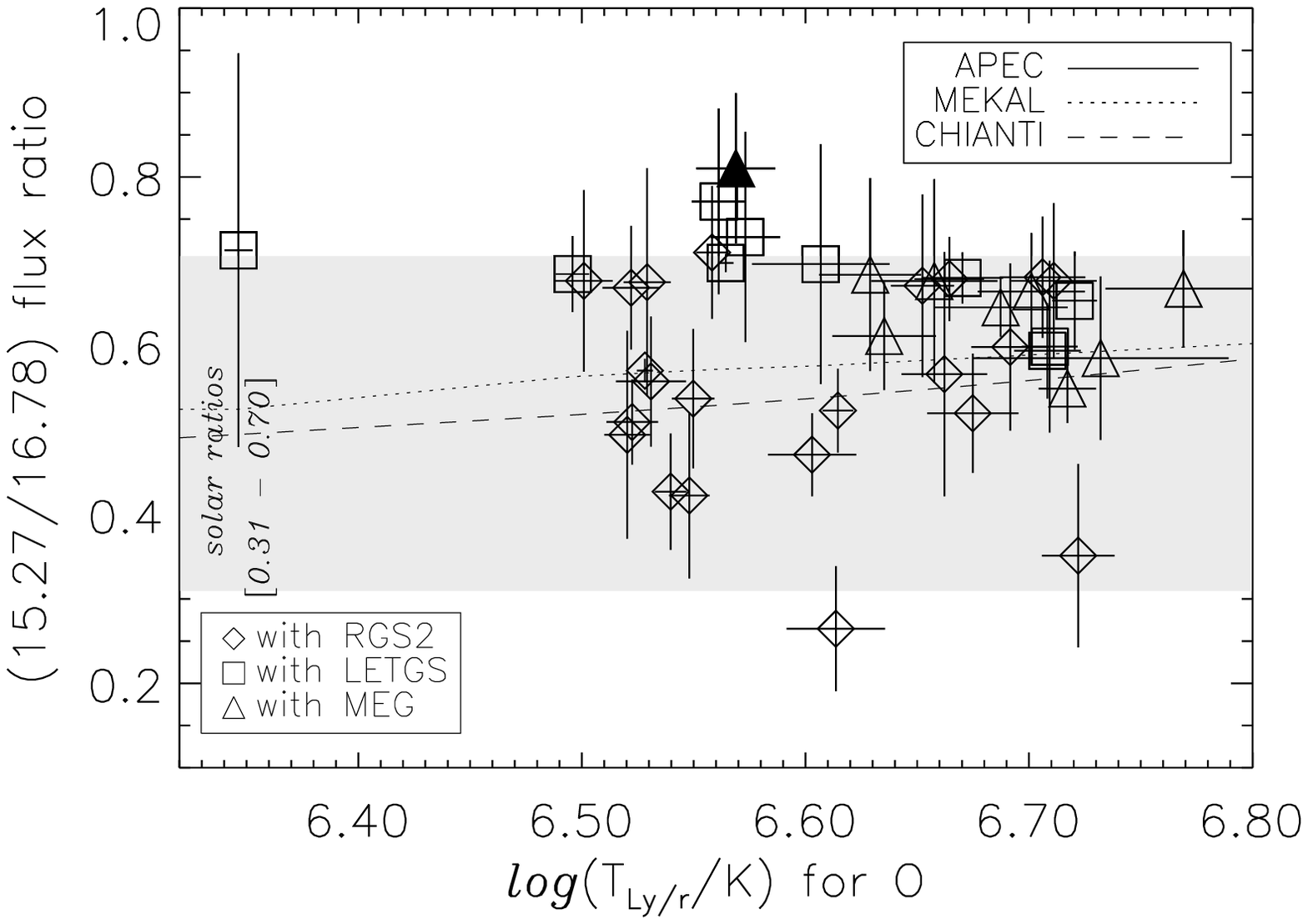}\includegraphics{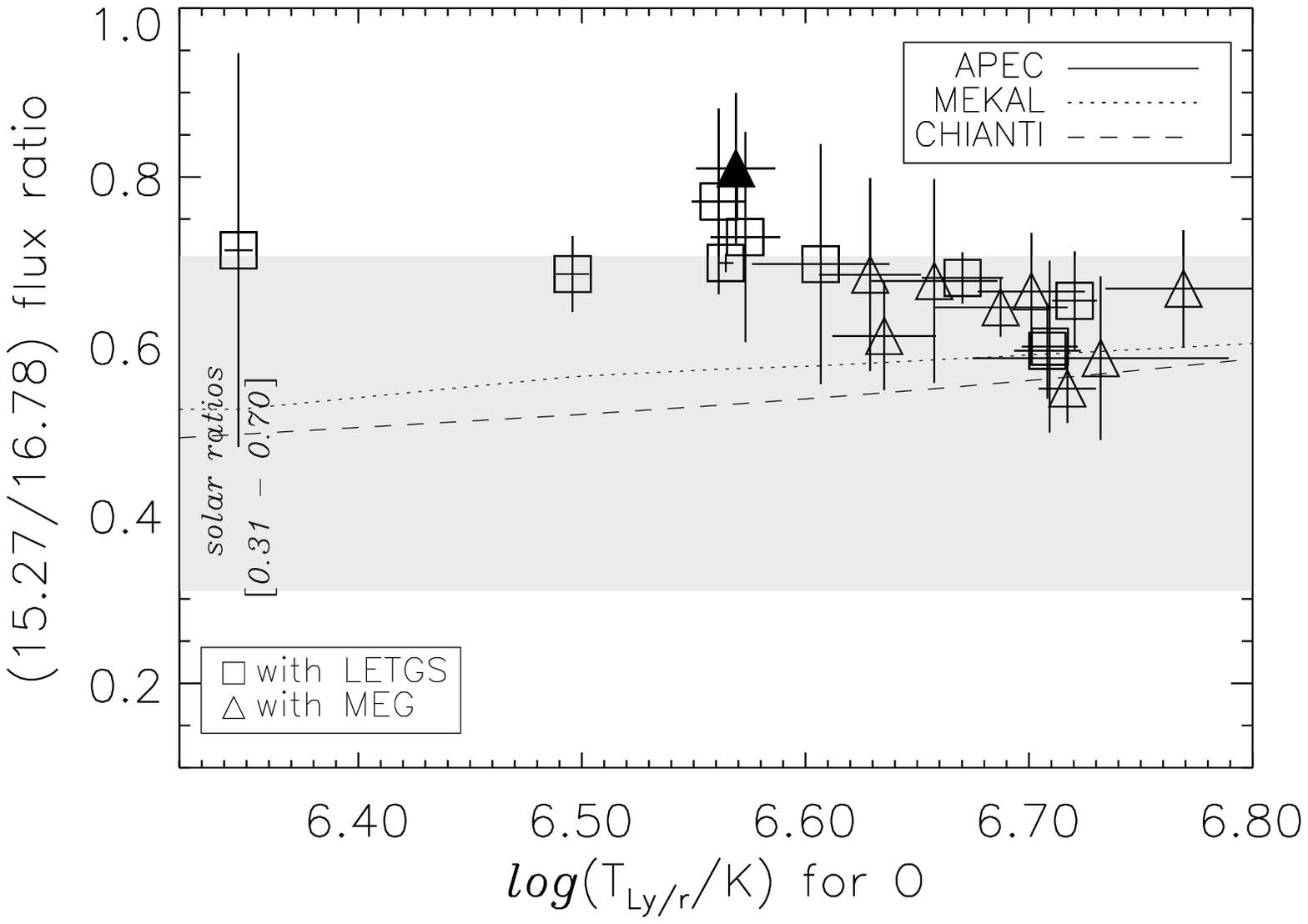}}
\caption{\label{fecheck}Ratio of Fe\,{\sc xvii} lines at 15.27\,\AA\ and
16.78\,\AA, which should not be sensitive to resonant scattering
(Right panel: only the MEG and LETG measurements).}
\end{figure*}

\begin{figure*}[!ht]
  \resizebox{\hsize}{!}{\includegraphics{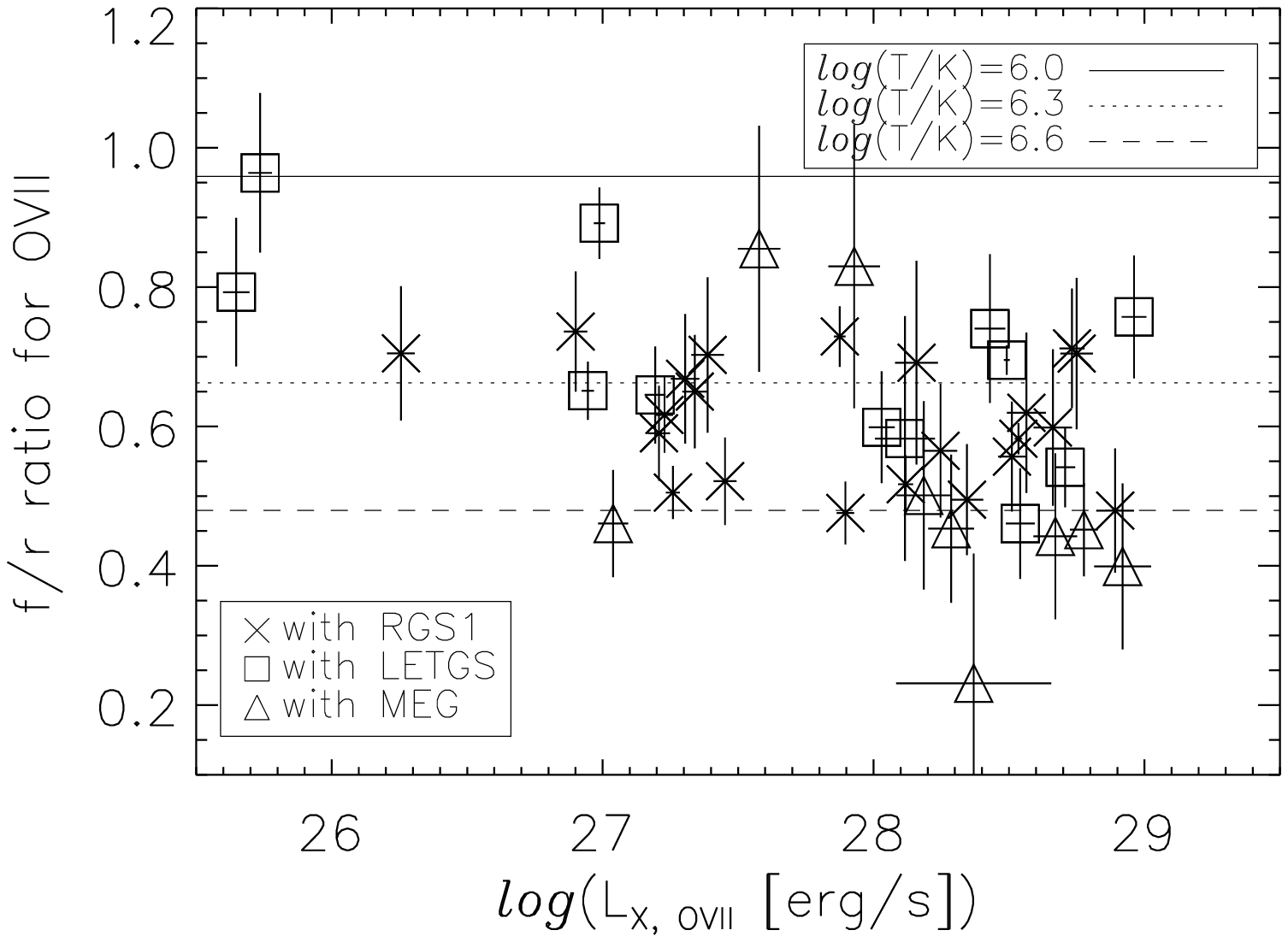}\includegraphics{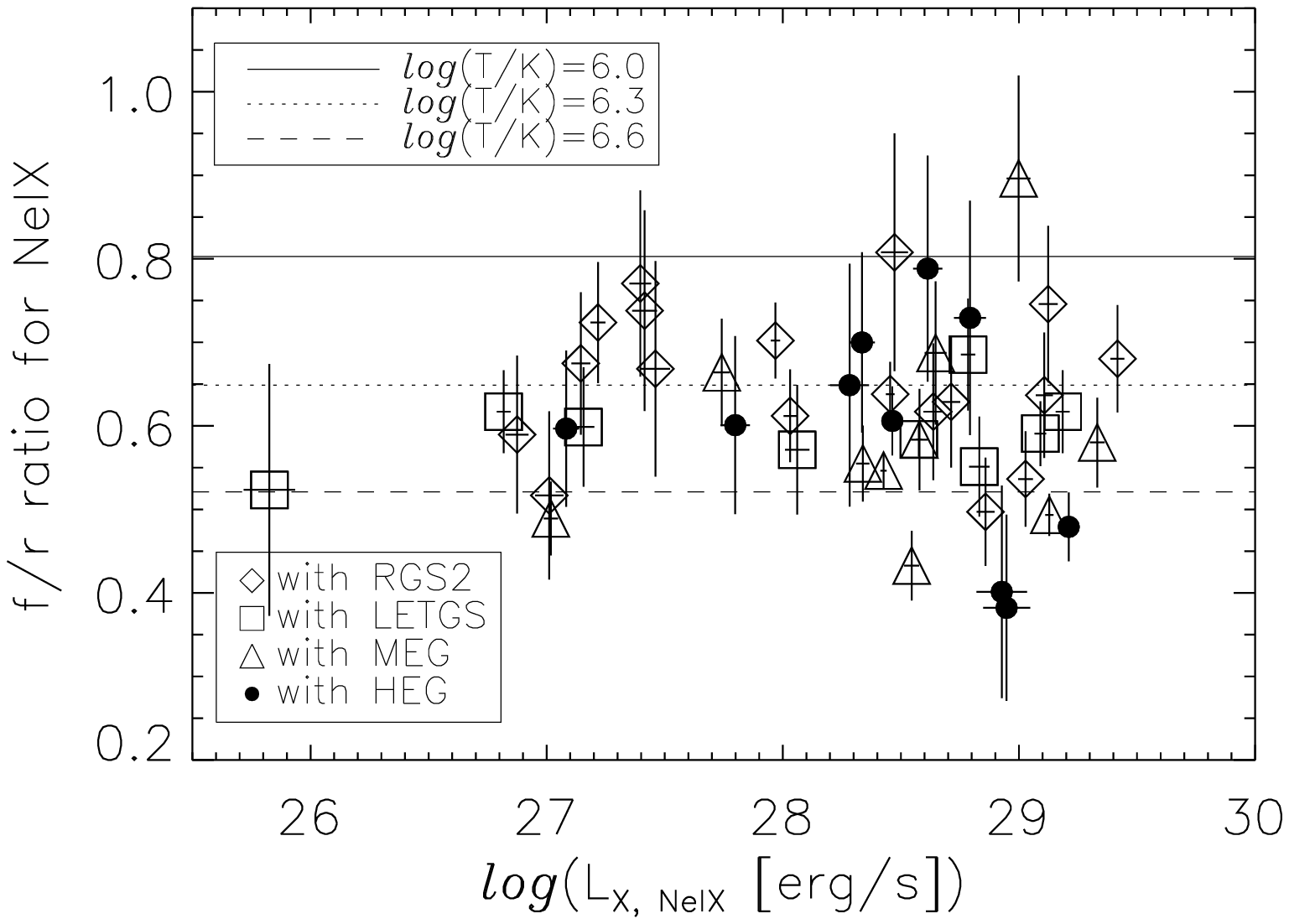}}
\caption{\label{Gratios}He-like f/r ratio for O\,{\sc vii} (left panel)
and for Ne\,{\sc ix} (right panel) versus the total He-like line luminosity
(r+i+f). Measurements with RGS1, RGS2, LETGS, MEG, and HEG
are compared with predictions from the APEC database for temperatures
log(T/K)=6.0, 6.3, and 6.6. assuming low densities.}
\end{figure*}

\subsection{Fe\,{\sc xvii} line ratios}
In Fig.~\ref{feratios} we plot the Fe\,{\sc xvii} line ratios of
15.27/15.03\,\AA\ lines and for 16.78/15.03\,\AA\ lines from Table~\ref{fe_cts}
versus O\,{\sc viii}/O\,{\sc vii} characteristic temperatures used as activity
indicators (listed in Table~\ref{fr_ox}). The horizontal lines represent
theoretical low-optical depth ratios as a function of temperature predicted by
interpolation from MEKAL, APEC, and Chianti, respectively. For the
16.78/15.03\,\AA\ line ratios we also included new theoretical predictions by
\cite{doron}, who account for dielectronic and radiative recombination from
Fe\,{\sc xviii}, inner-shell ionization from Fe\,{\sc xvi}, and resonant excitation
through doubly excited levels of Fe\,{\sc xvi} (3-ion model) in their calculations.
The model predictions lie significantly higher than the predictions from the
other databases, but their 1-ion model and their predictions for the
15.27/15.03\,\AA\ line ratio are consistent with the other predictions.\\
Comparing our measurements of the 15.27/15.03\,\AA\ line ratios with the
theoretical predictions the measured ratios are systematically higher than
predicted with no apparent correlation with temperature except possibly for the
coolest coronae in our sample where the ratios are highest. For the
16.78/15.03\,\AA\ ratios we find no correlation with temperature at all, but a
larger scatter with systematic deviations from the databases, although good
agreement with the predictions by \cite{doron} is seen.\\
In Fig.~\ref{fe_meg} we plot only the LETGS and MEG measurements, where the
scatter due to systematic and statistical uncertainties is much smaller. The
reason is that the RGS ratios suffer from systematic uncertainties in the
placement of the source continuum (due to broad line wings; cf. Sect.~\ref{bg})
and the HEG measurements have low signal to noise and have thus large statistical
uncertainties. In the left panel of Fig.~\ref{fe_meg} it can be seen that the
15.27/15.03\,\AA\ ratio is remarkably constant for all sources except
for the MEG measurement of EV Lac, which deviates considerably from all the
other MEG and LETGS measurements; this data point is marked by a filled
triangle. Both Fe\,{\sc xvii} ratios thus suggest that resonant scattering plays
a significant role for EV Lac, however, this high ratio cannot be confirmed in the
simultaneous HEG measurement nor in the RGS2 data. We show the two spectra obtained
with MEG and RGS2 for EV Lac in Fig.~\ref{evlac}, where the different ratios can
be recognized.

In order to compare our measurements with solar measurements we include the
solar measurements from \cite{saba99} in the form of shaded areas in
Figs.~\ref{feratios} and \ref{fe_meg}. They deduced significant optical
depths for the 15.03\,\AA\ resonance line by comparison with databases
available at that time. From the left panel of Fig.~\ref{feratios} it can be
seen that most of our measured ratios are located in the bottom part of the
shaded area, but only measurements for cooler coronae are really consistent
with solar measurements. For the 16.78/15.03\,\AA\ ratio we find solar
measurements significantly higher than all our results.\\
The calibration used for obtaining the solar line ratios
cannot be reconstructed, such that systematic uncertainties cannot be excluded
as the reason for the discrepancies. However, since ratios of very nearby lines
are calculated, only the relative calibration matters, which is always more
accurate than the absolute calibration for such nearby lines.
We point out that measurements for the Sun can also lead to different results
when specific regions in the solar corona are selected, while for the stars in
our sample only overall line fluxes can be obtained.

Contamination is always an issue that needs to be checked. We therefore
inspected the line flux ratios of the two low-$f$ lines at 15.27\,\AA\ and at
16.78\,\AA\ in Fig.~\ref{fecheck}, which should be independent of resonant
scattering effects. This ratio is consistent with both the solar measurements
and with the databases. Possible blending of the 15.27\,\AA\ line can explain
the enhanced ratios measured for the cooler coronae in the left panel of
Fig.~\ref{feratios}. Such enhancements can also be identified in
Fig.~\ref{fecheck}, but not for the 16.78/15.03\,\AA\ ratios.

\subsection{He-like line ratio f/r}
Another line ratio sensitive to resonance line scattering is the ratio f/r for
He-like ions, where f is the forbidden line $^3$S$_1\rightarrow^1$S$_0$ and r
is the resonance line $^1$P$_1\rightarrow^1$S$_0$. This ratio is also sensitive
to density and temperature. Interference with density effects is not severe
as \cite{ness_10} found low density limits for almost all stellar coronae.
In Fig.~\ref{Gratios} we plot the measured f/r
ratios for O\,{\sc vii} and Ne\,{\sc ix} versus the luminosity contained in all
three He-like lines of the respective ions, thus restricting the analysis to
only the plasma regions actually emitting in the respective lines. We over-plot
expected f/r ratios calculated from APEC for three temperatures log(T/K)=6.0,
6.3, and 6.6 assuming low densities. Good agreement between measurements
and predictions can be seen. The oxygen ratios seem to generally follow the
temperature trend suggested by the three theoretically predicted ratios,
decreasing with increasing degree of activity. For the Ne\,{\sc ix} f/r ratios
the scatter becomes larger for the more active stars, which must be
attributed to more severe blending by hotter Fe\,{\sc xix} lines; the blending
of Ne\,{\sc ix} by Fe\,{\sc xix} has been discussed by \cite{neix}, however,
for many stars the Fe lines blending the resonance and the forbidden lines are
relatively weak due to high Ne/Fe abundance ratios.

\section{Discussion}
\label{disc}

One of the major aims pursued by the analysis of coronal spectra is to understand
geometrical configurations of the coronal plasma. Opacity effects would make
the interpretation tremendously more complicated, because assumptions about
the geometrical configuration, which we want to study in the first place, would
have to be made in order to account for these effects. If resonant scattering
played an important role, one would naively expect that with an increasing
amount of plasma these effects would become more and more visible, thus the
more active stars should exhibit stronger effects on the line ratios
sensitive to resonant scattering. Therefore our analysis focuses on searching
for ratios of possibly optically thick resonance lines and optically thinner
lines to correlate with the degree of activity.\\
When drawing conclusions out of the measured ratios one has to keep in mind
that these ratios are not always determined by resonant scattering effects
alone, but might be obstructed by other effects as, e.g., line blending or
density effects. Possible temperature effects are considered quantitatively by
use of characteristic coronal temperatures derived from ratios of O\,{\sc viii}
and O\,{\sc vii} resonance lines.\\

\subsection{Stellar data}

\begin{figure}[!ht]
  \resizebox{\hsize}{!}{\includegraphics{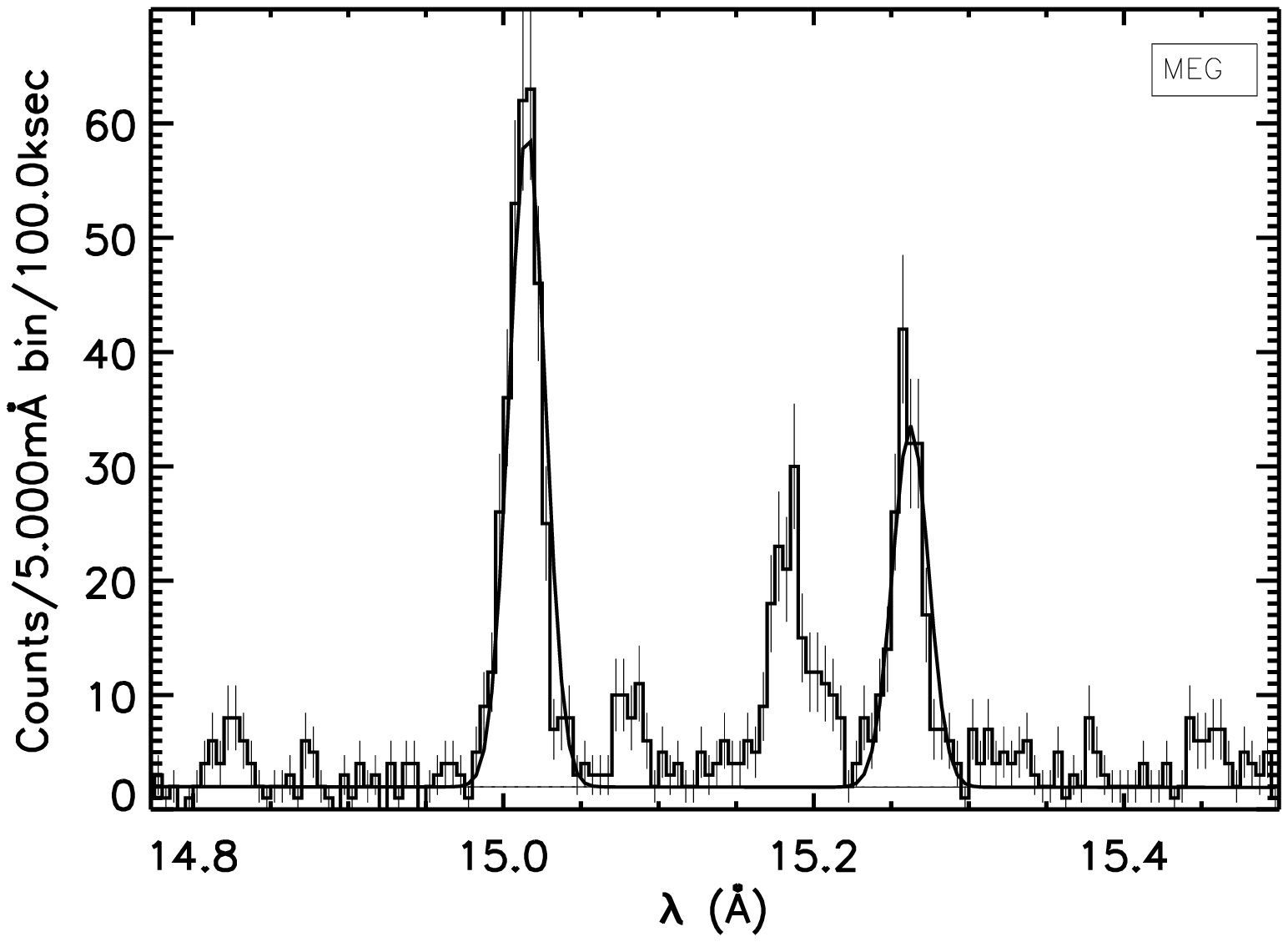}}
  \resizebox{\hsize}{!}{\includegraphics{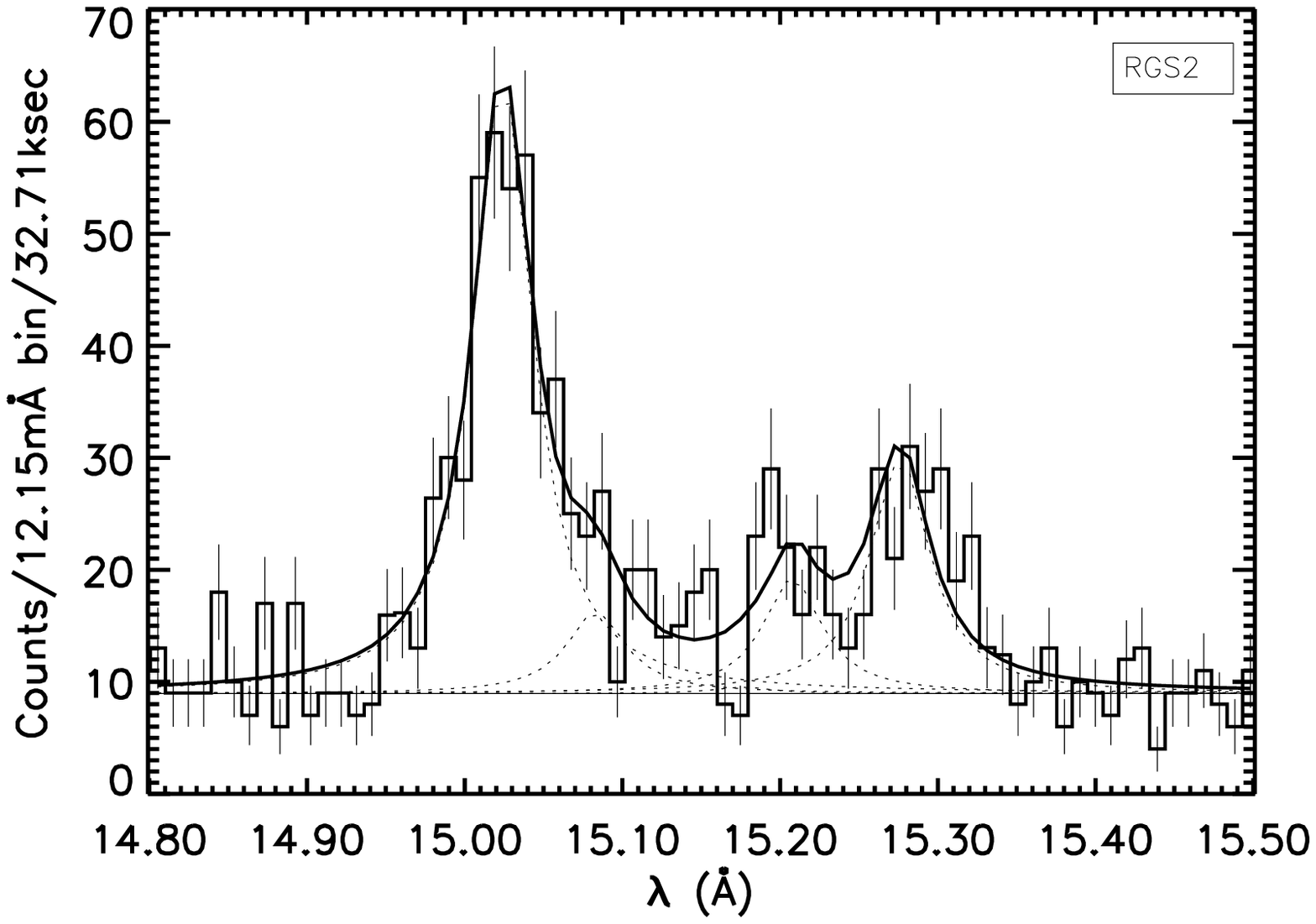}}
  \resizebox{\hsize}{!}{\includegraphics{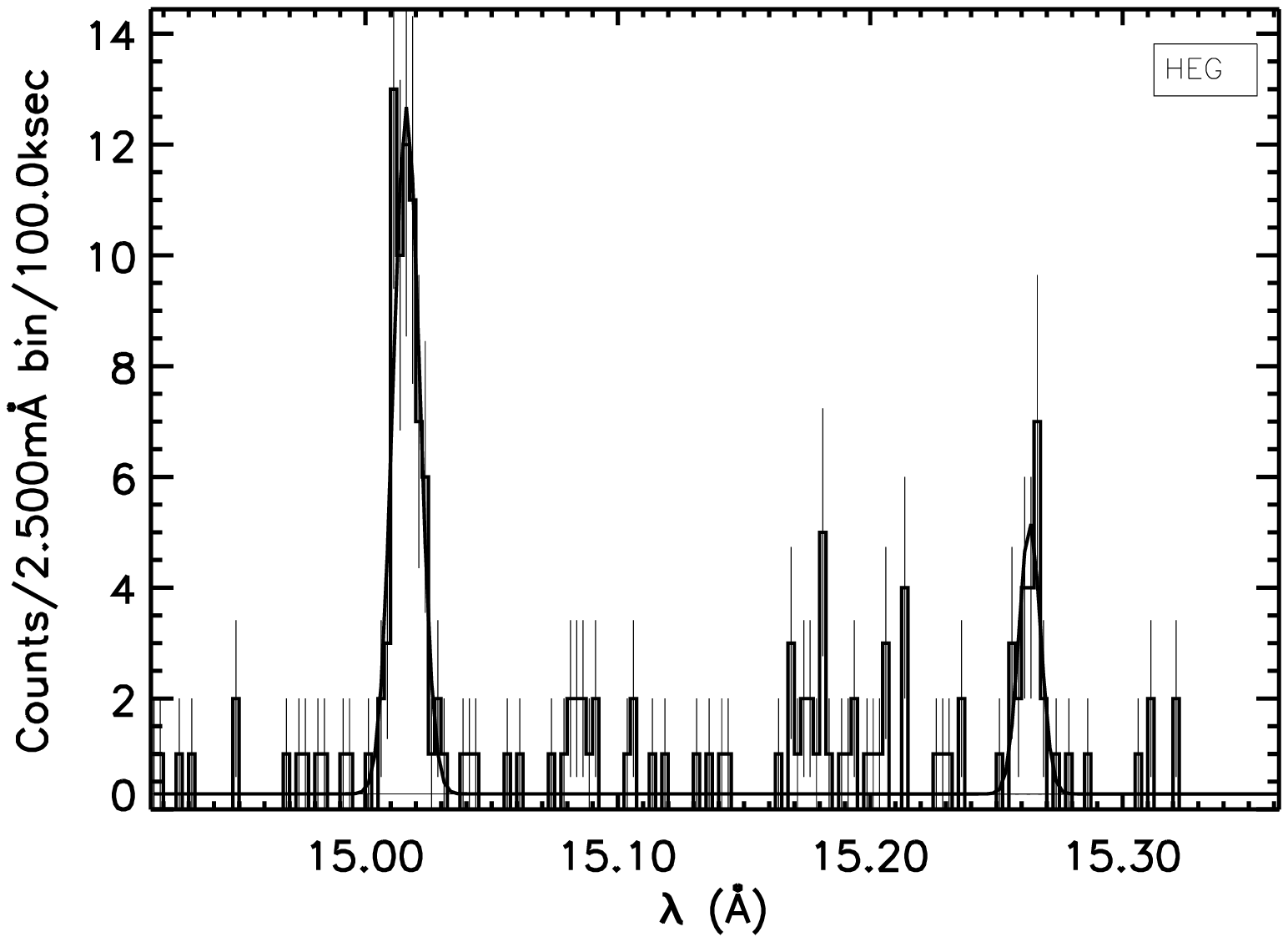}}
\caption{\label{evlac}Spectral region around 15\,\AA\ for EV Lac with
MEG (upper panel), with the RGS2 (middle panel), and with the HEG (bottom
panel). Significant differences in the 15.27\,\AA\ line compared with the
15.03\,\AA\ line can be recognized.}
\end{figure}

Our measured ratios of Fe\,{\sc xvii} 15.27/15.03\,\AA\ show a temperature
trend indicating that enhanced line ratios are found particularly for the
coolest coronae in our sample, but for the more active stars all measured line
ratios scatter around a constant value of about 0.38\,$\pm$\,0.07 with a
slight, but insignificant increasing trend towards higher temperatures.
All measured ratios are higher than predicted by the three data bases MEKAL,
APEC, and Chianti, but are consistent with laboratory measurements for low
optical depths obtained with EBIT. Many of the 16.78/15.03\,\AA\ line ratios
are discrepant with theoretical predictions and as a sample the ratios seem
generally to be discrepant with theory, but not with recent calculations by
\cite{doron} that include additional processes than pure collisional
excitation. No temperature trend can be seen in these data, and the
scatter is much larger than for the other ratio. This scatter cannot be
attributed to statistical uncertainties, because the 16.78\,\AA\ line is much
stronger than the 15.27\,\AA\ line. In Fig.~\ref{fe_meg} only the ratios with
the highest precision are plotted and still the 16.78\,\AA\ line seems more
problematic than the ratios with the 15.27\,\AA\ line. The 16.78/15.03\,\AA\
ratio is significantly more sensitive to resonant scattering than the
15.27/15.03\,\AA\ line ratio. The larger scatter could thus represent a variety
of resonant scattering processes.

The interpretation of opacity effects affecting only the 15.03\,\AA\
line (but not the 15.27\,\AA\ and the 16.78\,\AA\ lines) is certainly a
possible explanation for these deviations. From the 15.27/15.03\,\AA\ ratios this
would mean that opacity effects play a larger role for the inactive stars and
the Sun. However, such a temperature trend cannot be identified in the
16.78/15.03\,\AA\ line ratios lending support to the suspicion by
\cite{brown01} about blending of the Fe\,{\sc xvii} 15.27\,\AA\ line by
an Fe\,{\sc xvi} satellite line. However, the cited databases don't give
clear indications about the nature of such a blending line, so that no
clear identification can be given here. An Fe\,{\sc xvi} satellite would
disappear in the hotter coronae and leaves un-blended 15.27/15.03\,\AA\ line
ratios for these coronae. Such a trend should also be visible in
Fig.~\ref{fecheck}, but can only be recognized when concentrating on the LETGS
and MEG measurements. From Fig.~\ref{fecheck} we must conclude that the
blending scenario cannot explain all discrepancies, unless a similar blending
applies also to the 16.78\,\AA\ line.\\

Inspection of Fig.~\ref{fe_meg} (left panel) clearly shows that all
15.27/15.03\,\AA\ ratios measured with high confidence are systematically
enhanced above the predicted ratios, but no temperature trend can be seen,
neither in the data nor in the predictions. If taken at face value, these
deviations suggest that the opacities are significantly non-zero for all
stars, but also that optical depths are practically identical for all stars
given our heterogeneous sample. Alternatively, if none of the investigated
stellar coronae is optically thick, which would be quite surprising, the
deviations from the databases would then have to be explained by uncertainties
in the databases. For the 16.78/15.03\,\AA\ line ratio the databases
agree with each other, but when using more recent calculations by \cite{doron}
better agreement with our measurements can be seen. For the 15.27/15.03\,\AA\
ratio laboratory measurements disagree with the theoretical predictions. This
demonstrates that the inclusion of all kinds of side effects can change
theoretical predictions significantly. The ratios of the two low-$f$
Fe\,{\sc xvii} lines at 15.27\,\AA\ and 16.78\,\AA\ are plotted in
Fig.~\ref{fecheck} and agreement with theoretical predictions can be seen. From
this it is suggestive that problems in the databases might rather lie in
determining line fluxes for the 15.03\,\AA\ resonance line.\\

We marked the Fe\,{\sc xvii} line ratios measured with MEG for the flare star
EV Lac by filling its symbol assigned to the MEG (triangle), because in
Fig.\ref{fe_meg} this measurement is the only ratio significantly above the
otherwise flat trend for the 15.27/15.03\,\AA\ ratio. In Fig.~\ref{evlac} we
show the spectra of EV Lac obtained with MEG and RGS2 in order to demonstrate
that a significant difference can already be recognized by inspecting the
spectra. However, no event such as, e.g., marked flare activity during one of
the observations, can be associated with such a difference. In addition the HEG
spectrum shown in the bottom panel of Fig.~\ref{evlac} is rather
consistent with the RGS2 measurement although simultaneously observed with the
MEG.\\

Our second attempt to test for opacity effects and to probe possible other
emitting regions is the ratio f/r of the He-like ions O\,{\sc vii} and
Ne\,{\sc ix}. Although this ratio is also sensitive to density and
temperature, we find roughly the same f/r ratios for all stars in our sample,
definitely in agreement with plausible temperatures. No indication can be seen
suggesting opacity effects from these ratios. The dependence on temperature might,
however, be stronger than the sensitivity to resonant scattering effects. The
f/r ratios are therefore only useful in cases of resonant scattering effects
that outweigh the temperature sensitivity. For our measured ratios this means
that we can exclude strong resonant scattering effects, but weaker effects
could be hidden in the temperature trend.\\

\subsection{Comparison with solar measurements}

As described in Sect.~\ref{intro} the discussion about opacity effects in the
solar corona has been quite controversial. Since with the Sun only one
star is investigated our sample of 26 stellar coronae gives more
insight into trends or systematic effects. We focus on the Fe\,{\sc xvii}
line ratios, where solar measurements resulted in tempting evidence that the
15.03\,\AA\ line was significantly damped due to resonant scattering
\citep[e.g.,][]{schmelz97,saba99}. Discrepancies between theoretical
predictions and laboratory measurements made it difficult to identify the
measured ratios as pure resonant scattering effects. In addition the MEKAL
database has been upgraded since then and more refined databases have become
available.\\

We use the most recent databases and find that the discrepancies between
theoretical predictions and ratios for the Sun are still present. From the left
panel of Fig.~\ref{feratios} it can be seen that the 15.27/15.03\,\AA\ ratios
measured for the Sun
are consistent with the coolest coronae in our sample but not with the hotter
coronae and not with any of the more recent databases. A blending scenario for
the 15.27\,\AA\ line by Fe\,{\sc xvi} could explain the discrepancy and it
would be well consistent with the temperature trend found from our sample, but
it cannot be confirmed from the measured 15.27/16.78\,\AA\ line ratios.
With the solar data a blending scenario could not be identified, because the
temperatures encountered for the hotter coronae are never reached in the solar
corona and a blending Fe\,{\sc xvi} satellite line would thus never disappear.
In the right panel of Fig.~\ref{feratios} the solar measurements for the
16.78/15.03\,\AA\ line ratio are all systematically higher than all the ratios
from our sample as well as the predictions from the three databases MEKAL,
Chianti, and APEC. The predicted ratios from \cite{doron} are
consistent below log(T)$\sim$ 6.5, but most solar measurements are well above
these predictions as well. Note that the discrepancies between solar ratios
and our measurements are greater for the 16.78/15.03\,\AA\ ratio, which is
significantly more sensitive to resonant scattering effects than the
15.27/15.03\,\AA\ ratio. Systematic uncertainties in the calibration can of
course always lead to such deviations, since totally different instruments
were used to obtain the solar ratios. The calibration of our instruments seem
sufficiently well in order to produce similar results. If calibration error can
be excluded, a physical interpretation must be found for understanding why the
Sun should be the only star where opacity effects in the corona play a role.\\

\section{Summary and Conclusions}
\label{concl}

In the solar context opacity effects in X-ray lines have been discussed
controversially. In practice the strongest lines are used for the analysis
rather than weaker lines, but these lines are often resonance lines and are the
first candidates for opacity effects. We test for effects of resonant
scattering by measuring ratios of such resonance lines and forbidden lines with
significantly lower probabilities for such effects sampling many different
coronae.\\
The Fe\,{\sc xvii} 15.27/15.03\,\AA\ and 16.78/15.03\,\AA\ line ratios we measure
systematically higher than theoretical predictions, but for all kinds of different
coronae these deviations are strikingly similar. For the coolest coronae in our
sample we measure 15.27/15.03\,\AA\ ratios systematically higher ratios consistent
with solar measurements of the same ratio. This trend suggests blending of the
15.27\,\AA\ line. This trend can also be seen in the 15.27/16.78\,\AA\ ratios, but
only when ignoring the RGS2 measurements. In our large sample the 15.27/15.03\,\AA\
measurement for EV Lac deviates from the general trend, but an exceptional case
cannot be claimed because the simultaneous HEG observation is not consistent with
the MEG ratio. We interpret this measurement as a statistical outlier. For the
He-like f/r ratios for oxygen and neon all ratios can be explained by reasonable
temperatures, such that strong resonant scattering effects are ruled out. It cannot
be excluded, that weak resonant scattering effects are hidden in the large range of
ratios allowed for a reasonable temperature range.\\
Obviously the behavior with respect to resonant scattering is very similar for
all stars in our sample. Formally one could derive optical depths $\tau$ from
the measured deviations from the databases, however, similar but non-zero
optical depths for stars in all kinds of stellar activity are unlikely. We
therefore conclude that opacity effects should be considered as weak and
undetected and uncertainties in the databases could be a more plausible
explanation for the discrepancies. Since the 15.27/16.78\,\AA\ ratios are well
consistent with predictions from the databases we conclude that uncertainties
in the databases must lie in the 15.03\,\AA\ line.\\
The large discrepancies of our measurements with solar ratios are somewhat
puzzling. We doubt that opacity effects play a role only for the Sun. Also the
statistical argument cannot be applied in the way the EV Lac observation can be
treated, since many observations for the solar coronae exist. The high
15.27/15.03\,\AA\ could be explained by blending of the 15.27\,\AA\ line,
because the solar coronal plasma is in the right temperature range. Possibly
geometric effects might play a role. Observations of isolated emitting regions
on the solar surface might exclude resonant photons from the analysis that are
scattered out of not observed regions into the line of sight. Overall
measurements for stars collect all photons emitted towards the observer.\\

The methods we chose to investigate for resonant scattering effects in
coronal plasmas are commonly accepted to efficiently probe for these
effects. With the amount of data gathered with the new X-ray telescopes
we are convinced to operate on a sufficiently representative basis in order
to decide about optical thickness of coronal plasmas in general. As to
answering the question in our title we find deviations of measurements and
theoretical predictions that allow the conclusion of measurable resonant
scattering effects, however, we are not convinced that this conclusion is the
final answer. We did find systematic deviations of line ratios from optically
thin theoretical predictions, but we also found that theoretical predictions
can suffer from quite some uncertainties particularly when it comes down to
accounting for certain side effects. We also found striking similarities
between the ratios measured for all kinds of different coronae. From the
complicated geometrical configurations expected for coronal plasma optical
depths are unlikely to be so similar for inactive, intermediately active, and
most active coronae. The amount of emitting plasma being up to four
orders of magnitude different in X-ray luminosity raises the expectation that
optical depths will be much larger for active stars than for inactive stars. The
only scenario that we find plausible on the background of such similarities is
that resonant scattering effects are all in the same way not detectable.\\

The detection of resonant scattering for the Sun seems to be a different story.
We attribute these differences to some resonant scattering effects possibly
always taking place. For the Sun these effects are better detectable when
focusing on selected regions, while for stellar observations these effects are
balanced out by observing globally. It must be pointed out that no measurement
for the Sun has been reported describing any kind of ''negative resonant
scattering'' that could balance out hypothetical global observations for the
Sun.\\

What does it mean when we conclude resonant scattering to be taking place
without being detectable for stellar coronae? Practically the analysis
of stellar coronal emission always refers to globally averaged statements, not
only for the aspect of resonant scattering. When analysing resonance lines
the non-detectability of existent resonant scattering effects means that
statements derived out of, e.g., ratios of resonance lines, made on a global
basis are still valid. It has to be kept in mind that no statements can be
made for individual emitting regions, but that only average conclusions about
all kinds of different emitting regions can be drawn, and on this level a
balance of resonant scattering effects is equivalent with negligible resonant
scattering effects.

\begin{acknowledgements}
This work is based on observations obtained with XMM-Newton, an ESA science
mission with instruments and contributions directly funded by ESA Member
States and the USA (NASA).\\
J.-U.N. acknowledges support from DLR under 50OR0105. MA and MG acknowledge
support from the Swiss National Science Foundation
(fellowship 81EZ-67388 and grant 2000-058827).
The SRON National Institute for Space Research is supported financially by
NWO. We thank our referee Dr. F.P.Keenan.

\end{acknowledgements}

\end{document}

%% file: tab1.tex
\begin{table*}
\begin{flushleft}
\renewcommand{\arraystretch}{1.2}
\caption{\label{sample}Properties of observed sample of 26 stars with 45 analysed spectra}
{\scriptsize
\begin{tabular}{p{.4cm}rcccc|ccccccc}
 star& Spectr.&dist.&\multicolumn{3}{c|}{exposure time [ks]} & \multicolumn{7}{c}{L$_X$(10$^{28}$\,erg/s)}\\
 & type& [pc] & RGS1,2 & LETG & HEG/& \multicolumn{2}{c}{RGS1} & \multicolumn{2}{c}{RGS2} & LETG & MEG & HEG\\
 & &&&&\ MEG&1$^{\rm st}$order&2$^{\rm nd}$order&1$^{\rm st}$order&2$^{\rm nd}$order&&\\
       47\,Cas$^a$&             F0.0Vn&   33.56&   50.66&    --&    --&  112.88&  161.85&  184.34&  122.27&    --&    --&    --\\
       AB\,Dor$^b$&           K1.0IIIp&   14.94&   58.87&    --&   52.33&   70.87&   50.96&   65.81&   44.72&    --&   73.87&   58.21\\
  $\alpha$\,Cen\,A&                G2V&    1.34&    --&   81.50&    --&    --&    --&    --&    --&    0.08&    --&    --\\
  $\alpha$\,Cen\,B&                K0V&    1.34&    7.04&   81.50&    --&    --&    --&    --&    --&    0.07&    --&    --\\
           AD\,Leo&              M3.5V&    4.70&   36.25&   48.50&    --&    3.50&    3.01&    3.07&    2.03&    3.93&    --&    --\\
             Algol&      B8.0V/K2.0III&   28.00&   52.66&   81.40&   51.73&  283.73&    --&    --&    --&  944.99&  673.63&  557.72\\
           AR\,Lac&             G2.0IV&   42.03&   32.10&    --&   32.09&  627.14&  612.06&  730.55&  464.23&    --&  514.49&  415.12\\
           AT\,Mic&               M4.4&   10.22&   28.19&    --&    --&   15.47&   14.44&   20.49&   11.46&    --&    --&    --\\
       AU\,Mic$^a$&               M0.0&    9.94&   55.74&    --&   58.81&   12.43&   14.42&   18.31&   10.53&    --&   11.52&    8.55\\
      $\beta$\,Cet&            K0.0III&   29.38&   13.02&  108.04&   86.06&  198.67&  320.16&  328.04&  256.10&  699.24&  253.19&  227.48\\
       Capella$^b$&           G5.0IIIe&   12.94&   52.92&  218.50&  154.68&  145.38&  157.22&  168.06&  132.10&  186.56&  153.08&  127.43\\
     $\chi^1$\,Ori&              G0.0V&    8.66&   29.96&    --&    --&    6.21&    6.14&    4.95&    4.09&    --&    --&    --\\
           EK\,Dra&               F8.0&   33.94&   34.31&   67.24&    --&   61.25&   72.78&   97.76&   54.54&   85.01&    --&    --\\
   $\epsilon$\,Eri&              K2.0V&    3.22&   13.34&  108.00&    --&    0.55&    0.55&    --&    0.53&    1.53&    --&    --\\
       EQ\,Peg$^b$&               M3.5&    6.25&   15.52&    --&    --&    4.31&    3.48&    3.81&    2.80&    --&    --&    --\\
           EV\,Lac&               M3.5&    5.05&   32.71&    --&  100.06&    3.58&    2.97&    4.03&    2.11&    --&    2.86&    2.19\\
          HR\,1099&              G9.0V&   28.97&   26.27&   97.50&   94.68&  432.49&  384.98&  488.85&  313.58&  973.74& 1000.85&  798.38\\
 $\kappa$\,Cet$^a$&           G5.0Vvar&    9.16&   39.95&    --&    --&    5.61&    5.27&    2.79&    3.44&    --&    --&    --\\
    $\lambda$\,And&            G8.0III&   25.81&   31.83&    --&   81.91&  249.61&  238.61&  295.03&  188.31&    --&  198.33&  121.17\\
      $\pi^1$\,UMa&             G1.5Vb&   14.27&   52.90&    --&    --&    2.56&    6.19&    5.13&    4.14&    7.29&    --&    --\\
           Procyon&           F5.0IV-V&    3.50&    --&  140.70&    --&    --&    --&    --&    --&    0.49&    --&    --\\
     $\sigma$\,CrB&               G0.0&   21.70&   19.31&    --&    --&  322.59&  338.72&  307.83&  312.74&    --&    --&    --\\
           UX\,Ari&             G5.0IV&   50.23&   30.91&  112.76&   48.47&  953.03&  873.75& 1059.84&  691.82& 1306.36&  804.46&  502.40\\
           VY\,Ari&               K0.0&   43.99&   33.82&    --&    --&  366.79&  352.07&  460.52&  266.27&    --&    --&    --\\
           YY\,Gem&              M0.5V&   15.80&    --&   59.00&    --&    --&    --&    --&    --&   37.12&    --&    --\\
       YZ\,CMi$^a$&            M4.5V:e&    5.93&   27.24&    --&    --&    3.30&    1.59&    3.07&    1.94&    --&    --&    --\\
\hline
\end{tabular}
\\
$^a$RGS spectra with 95\,\% extraction regions\\
$^b$full RGS1 range available
}
\renewcommand{\arraystretch}{1}
\end{flushleft}
\end{table*}

%% file: tab2.tex
\begin{table*}[!ht]
\begin{flushleft}
\renewcommand{\arraystretch}{1.2}
\caption{\label{fe_cts}Measured line counts for Fe\,{\sc xvii} lines with 1\,$\sigma$ errors}
{\scriptsize
\begin{tabular}{p{.9cm}p{.7cm}ccc|ccc}
&&\multicolumn{3}{c|}{Fe\,{\sc xvii}}&\multicolumn{3}{c}{Ratios}\\
 star &Instr. & 15.03\,\AA & 15.27\,\AA & 16.78\,\AA & 15.27/15.03\,\AA& 16.78/15.03\,\AA & 15.27/16.78\,\AA\\
         47\,Cas&  RGS2&  506.28\,$\pm$\,32.80&  215.94\,$\pm$\,25.97&  419.17\,$\pm$\,26.64&     0.40\,$\pm$\,0.05&     0.77\,$\pm$\,0.07&     0.52\,$\pm$\,0.07\\
         AB\,Dor&  RGS2&  1132.6\,$\pm$\,48.68&  464.77\,$\pm$\,39.14&  905.31\,$\pm$\,40.05&     0.39\,$\pm$\,0.03&     0.74\,$\pm$\,0.04&     0.52\,$\pm$\,0.05\\
 &   MEG&  538.89\,$\pm$\,24.37&  203.40\,$\pm$\,15.96&  241.83\,$\pm$\,16.35&     0.39\,$\pm$\,0.03&     0.59\,$\pm$\,0.04&     0.66\,$\pm$\,0.06\\
 &   HEG&  91.010\,$\pm$\,9.892&  37.030\,$\pm$\,6.595&  18.580\,$\pm$\,4.359&     0.43\,$\pm$\,0.09&     0.62\,$\pm$\,0.16&     0.70\,$\pm$\,0.20\\
         AD\,Leo&  RGS2&  429.66\,$\pm$\,29.84&  159.63\,$\pm$\,23.61&  380.81\,$\pm$\,24.73&     0.35\,$\pm$\,0.05&     0.82\,$\pm$\,0.07&     0.42\,$\pm$\,0.06\\
 &  LETG&  313.64\,$\pm$\,20.17&  133.11\,$\pm$\,15.45&  189.05\,$\pm$\,15.70&     0.41\,$\pm$\,0.05&     0.53\,$\pm$\,0.05&     0.77\,$\pm$\,0.11\\
           Algol&  RGS2&  1028.4\,$\pm$\,49.94&  457.75\,$\pm$\,40.60&  684.78\,$\pm$\,39.29&     0.42\,$\pm$\,0.04&     0.62\,$\pm$\,0.04&     0.68\,$\pm$\,0.07\\
 &  LETG&  1134.5\,$\pm$\,42.19&  378.77\,$\pm$\,31.01&  698.16\,$\pm$\,33.95&     0.32\,$\pm$\,0.02&     0.54\,$\pm$\,0.03&     0.59\,$\pm$\,0.05\\
 &   MEG&  621.70\,$\pm$\,27.55&  226.82\,$\pm$\,18.68&  294.40\,$\pm$\,19.00&     0.38\,$\pm$\,0.03&     0.62\,$\pm$\,0.04&     0.61\,$\pm$\,0.06\\
 &   HEG&  104.25\,$\pm$\,10.66&  39.750\,$\pm$\,6.837&  13.140\,$\pm$\,3.780&     0.41\,$\pm$\,0.08&     0.38\,$\pm$\,0.11&     1.07\,$\pm$\,0.35\\
         AR\,Lac&  RGS2&  582.06\,$\pm$\,35.42&  259.47\,$\pm$\,29.81&  390.62\,$\pm$\,28.58&     0.42\,$\pm$\,0.05&     0.62\,$\pm$\,0.06&     0.67\,$\pm$\,0.09\\
 &   MEG&  218.91\,$\pm$\,15.85&  79.000\,$\pm$\,10.28&  107.15\,$\pm$\,10.98&     0.37\,$\pm$\,0.05&     0.64\,$\pm$\,0.08&     0.58\,$\pm$\,0.09\\
         AT\,Mic&  RGS2&  294.34\,$\pm$\,24.58&  88.730\,$\pm$\,18.89&  213.98\,$\pm$\,19.77&     0.28\,$\pm$\,0.06&     0.68\,$\pm$\,0.08&     0.42\,$\pm$\,0.09\\
         AU\,Mic&  RGS2&  857.76\,$\pm$\,40.54&  336.14\,$\pm$\,31.18&  477.49\,$\pm$\,28.79&     0.37\,$\pm$\,0.03&     0.52\,$\pm$\,0.04&     0.71\,$\pm$\,0.07\\
 &   MEG&  217.18\,$\pm$\,15.51&  78.240\,$\pm$\,9.768&  90.760\,$\pm$\,10.07&     0.37\,$\pm$\,0.05&     0.55\,$\pm$\,0.07&     0.68\,$\pm$\,0.11\\
 &   HEG&  28.080\,$\pm$\,5.613&  7.6600\,$\pm$\,3.173&  6.9400\,$\pm$\,2.646&     0.29\,$\pm$\,0.13&     0.75\,$\pm$\,0.32&     0.39\,$\pm$\,0.22\\
    $\beta$\,Cet&  RGS2&  752.18\,$\pm$\,40.03&  286.31\,$\pm$\,26.32&  619.54\,$\pm$\,30.46&     0.36\,$\pm$\,0.03&     0.77\,$\pm$\,0.05&     0.47\,$\pm$\,0.04\\
 &  LETG&  3718.0\,$\pm$\,70.06&  1248.1\,$\pm$\,45.71&  2008.5\,$\pm$\,50.64&     0.32\,$\pm$\,0.01&     0.48\,$\pm$\,0.01&     0.68\,$\pm$\,0.03\\
 &   MEG&  1954.0\,$\pm$\,45.23&  672.90\,$\pm$\,27.24&  827.40\,$\pm$\,29.22&     0.36\,$\pm$\,0.01&     0.56\,$\pm$\,0.02&     0.64\,$\pm$\,0.03\\
 &   HEG&  298.67\,$\pm$\,17.44&  108.19\,$\pm$\,10.61&  50.690\,$\pm$\,7.141&     0.39\,$\pm$\,0.04&     0.51\,$\pm$\,0.07&     0.75\,$\pm$\,0.13\\
         Capella&  RGS2&  11967.\,$\pm$\,145.8&  5284.1\,$\pm$\,110.5&  9436.1\,$\pm$\,117.3&     0.42\,$\pm$\,0.01&     0.73\,$\pm$\,0.01&     0.57\,$\pm$\,0.01\\
 &  LETG&  22759.\,$\pm$\,165.8&  8940.5\,$\pm$\,111.5&  14024.\,$\pm$\,124.6&     0.38\,$\pm$\,0.01&     0.54\,$\pm$\,0.01&     0.69\,$\pm$\,0.01\\
 &   MEG&  14047.\,$\pm$\,120.8&  5343.9\,$\pm$\,76.05&  6629.4\,$\pm$\,82.34&     0.40\,$\pm$\,0.01&     0.62\,$\pm$\,0.01&     0.64\,$\pm$\,0.01\\
 &   HEG&  2712.3\,$\pm$\,52.65&  843.36\,$\pm$\,29.72&  503.93\,$\pm$\,22.52&     0.33\,$\pm$\,0.01&     0.56\,$\pm$\,0.02&     0.59\,$\pm$\,0.03\\
   $\chi^1$\,Ori&  RGS2&  549.96\,$\pm$\,30.83&  265.64\,$\pm$\,24.36&  404.77\,$\pm$\,24.20&     0.46\,$\pm$\,0.04&     0.68\,$\pm$\,0.05&     0.66\,$\pm$\,0.07\\
         EK\,Dra&  RGS2&  280.71\,$\pm$\,22.36&  54.790\,$\pm$\,14.60&  210.94\,$\pm$\,17.96&     0.18\,$\pm$\,0.05&     0.70\,$\pm$\,0.08&     0.26\,$\pm$\,0.07\\
 &  LETG&  204.32\,$\pm$\,16.49&  65.458\,$\pm$\,10.53&  102.86\,$\pm$\,12.91&     0.31\,$\pm$\,0.05&     0.44\,$\pm$\,0.06&     0.69\,$\pm$\,0.14\\
 $\epsilon$\,Eri&  RGS2&  230.74\,$\pm$\,19.67&  123.29\,$\pm$\,16.27&  185.60\,$\pm$\,16.44&     0.50\,$\pm$\,0.08&     0.75\,$\pm$\,0.09&     0.67\,$\pm$\,0.10\\
 &  LETG&  1054.9\,$\pm$\,34.98&  462.69\,$\pm$\,24.38&  739.58\,$\pm$\,28.96&     0.42\,$\pm$\,0.02&     0.62\,$\pm$\,0.03&     0.68\,$\pm$\,0.04\\
         EQ\,Peg&  RGS2&  134.74\,$\pm$\,16.70&  59.830\,$\pm$\,13.14&  123.28\,$\pm$\,14.43&     0.42\,$\pm$\,0.10&     0.85\,$\pm$\,0.14&     0.49\,$\pm$\,0.12\\
         EV\,Lac&  RGS2&  403.93\,$\pm$\,28.68&  154.06\,$\pm$\,20.51&  292.15\,$\pm$\,22.44&     0.36\,$\pm$\,0.05&     0.67\,$\pm$\,0.07&     0.53\,$\pm$\,0.08\\
 &   MEG&  333.57\,$\pm$\,18.93&  183.79\,$\pm$\,14.40&  180.05\,$\pm$\,13.84&     0.57\,$\pm$\,0.05&     0.71\,$\pm$\,0.06&     0.81\,$\pm$\,0.08\\
 &   HEG&  62.580\,$\pm$\,8.187&  20.190\,$\pm$\,4.849&  10.800\,$\pm$\,3.313&     0.34\,$\pm$\,0.09&     0.52\,$\pm$\,0.17&     0.66\,$\pm$\,0.25\\
        HR\,1099&  RGS2&  553.53\,$\pm$\,37.37&  222.57\,$\pm$\,30.33&  337.93\,$\pm$\,29.10&     0.38\,$\pm$\,0.05&     0.57\,$\pm$\,0.06&     0.67\,$\pm$\,0.10\\
 &  LETG&  1204.3\,$\pm$\,42.48&  434.95\,$\pm$\,31.77&  728.82\,$\pm$\,36.98&     0.35\,$\pm$\,0.02&     0.54\,$\pm$\,0.03&     0.65\,$\pm$\,0.05\\
 &   MEG&  1227.7\,$\pm$\,39.05&  445.89\,$\pm$\,26.59&  644.34\,$\pm$\,28.31&     0.38\,$\pm$\,0.02&     0.69\,$\pm$\,0.03&     0.54\,$\pm$\,0.04\\
 &   HEG&  240.44\,$\pm$\,16.57&  72.760\,$\pm$\,10.13&  43.570\,$\pm$\,7.306&     0.32\,$\pm$\,0.05&     0.55\,$\pm$\,0.10&     0.59\,$\pm$\,0.12\\
   $\kappa$\,Cet&  RGS2&  550.57\,$\pm$\,31.34&  312.80\,$\pm$\,27.04&  619.60\,$\pm$\,29.86&     0.53\,$\pm$\,0.05&     1.05\,$\pm$\,0.07&     0.51\,$\pm$\,0.05\\
  $\lambda$\,And&  RGS2&  478.78\,$\pm$\,33.93&  197.61\,$\pm$\,28.29&  336.44\,$\pm$\,28.02&     0.39\,$\pm$\,0.06&     0.65\,$\pm$\,0.07&     0.59\,$\pm$\,0.09\\
 &   MEG&  558.06\,$\pm$\,25.05&  201.97\,$\pm$\,16.07&  241.32\,$\pm$\,16.13&     0.38\,$\pm$\,0.03&     0.57\,$\pm$\,0.04&     0.66\,$\pm$\,0.06\\
 &   HEG&  58.620\,$\pm$\,7.790&  23.100\,$\pm$\,4.943&                   -- &     0.42\,$\pm$\,0.10&                   -- &      --\\
    $\pi^1$\,UMa&  RGS2&  450.19\,$\pm$\,27.04&  176.91\,$\pm$\,21.40&  323.24\,$\pm$\,21.57&     0.37\,$\pm$\,0.05&     0.67\,$\pm$\,0.06&     0.55\,$\pm$\,0.07\\
         Procyon&  LETG&  101.56\,$\pm$\,13.37&  38.450\,$\pm$\,10.07&  59.030\,$\pm$\,11.56&     0.37\,$\pm$\,0.10&     0.51\,$\pm$\,0.12&     0.71\,$\pm$\,0.23\\
   $\sigma$\,CrB&  RGS2&  1626.6\,$\pm$\,56.60&  673.04\,$\pm$\,41.30&  1009.9\,$\pm$\,40.78&     0.39\,$\pm$\,0.02&     0.58\,$\pm$\,0.03&     0.67\,$\pm$\,0.05\\
         UX\,Ari&  RGS2&  220.17\,$\pm$\,27.35&  73.050\,$\pm$\,20.96&  211.92\,$\pm$\,24.48&     0.31\,$\pm$\,0.09&     0.90\,$\pm$\,0.15&     0.35\,$\pm$\,0.10\\
 &  LETG&  497.87\,$\pm$\,29.79&  151.85\,$\pm$\,22.32&  277.58\,$\pm$\,23.61&     0.29\,$\pm$\,0.04&     0.49\,$\pm$\,0.05&     0.59\,$\pm$\,0.10\\
 &   MEG&  219.83\,$\pm$\,16.26&  79.050\,$\pm$\,10.85&  92.730\,$\pm$\,10.55&     0.37\,$\pm$\,0.05&     0.55\,$\pm$\,0.07&     0.67\,$\pm$\,0.12\\
         VY\,Ari&  RGS2&  291.14\,$\pm$\,26.33&  90.040\,$\pm$\,20.14&  161.97\,$\pm$\,19.96&     0.29\,$\pm$\,0.07&     0.52\,$\pm$\,0.08&     0.56\,$\pm$\,0.14\\
         YY\,Gem&  LETG&  252.48\,$\pm$\,18.17&  92.900\,$\pm$\,12.66&  139.56\,$\pm$\,14.36&     0.35\,$\pm$\,0.05&     0.49\,$\pm$\,0.06&     0.72\,$\pm$\,0.12\\
         YZ\,CMi&  RGS2&  163.75\,$\pm$\,18.27&  97.110\,$\pm$\,16.44&  145.15\,$\pm$\,15.44&     0.56\,$\pm$\,0.11&     0.82\,$\pm$\,0.12&     0.67\,$\pm$\,0.13\\
\hline
\end{tabular}
}
\renewcommand{\arraystretch}{1}
\end{flushleft}
\end{table*}

%% file: tab3.tex
\begin{table*}[!ht]
\begin{flushleft}
\renewcommand{\arraystretch}{1.2}
\caption{\label{fr_ox}Measured line counts for O\,{\sc vii} resonance (r) and
forbidden (f) lines and O\,{\sc viii} (Ly$_\alpha$), O\,{\sc vii} luminosities
L$_{\rm O \mathsc{vii}}$, and O\,{\sc viii}/O\,{\sc vii} temperatures.}
{\scriptsize
\begin{tabular}{p{.9cm}p{.8cm}cccccc}
 star & Instr.& r & f & f/r & log(L$_{\rm O \mathsc{vii}}$)$^a$ & Ly & T$^b$/MK\\
         47\,Cas&  RGS1&  186.55\,$\pm$\,18.08&  105.41\,$\pm$\,14.67&    0.565\,$\pm$\,0.09&    28.25\,$\pm$\,0.06&  1275.7\,$\pm$\,39.89&     4.73\,$\pm$\,0.22\\
         AB\,Dor&  RGS1&  788.24\,$\pm$\,35.68&  407.50\,$\pm$\,27.21&    0.517\,$\pm$\,0.04&    28.12\,$\pm$\,0.02&  3760.9\,$\pm$\,68.54&     4.12\,$\pm$\,0.06\\
 &   MEG&  44.570\,$\pm$\,7.304&  36.980\,$\pm$\,6.752&    0.829\,$\pm$\,0.20&    27.93\,$\pm$\,0.09&  814.23\,$\pm$\,28.94&     5.88\,$\pm$\,0.49\\
$\alpha$\,Cen\,A&  LETG&  138.26\,$\pm$\,12.31&  109.60\,$\pm$\,11.06&    0.792\,$\pm$\,0.10&    25.65\,$\pm$\,0.04&  60.890\,$\pm$\,8.748&     1.88\,$\pm$\,0.08\\
$\alpha$\,Cen\,B&  RGS1&  157.81\,$\pm$\,13.96&  111.23\,$\pm$\,11.60&    0.704\,$\pm$\,0.09&    26.26\,$\pm$\,0.05&  188.22\,$\pm$\,14.73&     2.60\,$\pm$\,0.07\\
 &  LETG&  160.32\,$\pm$\,13.35&  154.54\,$\pm$\,13.06&    0.963\,$\pm$\,0.11&    25.74\,$\pm$\,0.04&  130.31\,$\pm$\,12.45&     2.16\,$\pm$\,0.06\\
         AD\,Leo&  RGS1&  705.13\,$\pm$\,30.13&  356.13\,$\pm$\,22.02&    0.505\,$\pm$\,0.03&    27.26\,$\pm$\,0.02&  2132.8\,$\pm$\,49.62&     3.47\,$\pm$\,0.05\\
 &  LETG&  263.97\,$\pm$\,17.47&  170.31\,$\pm$\,14.52&    0.645\,$\pm$\,0.07&    27.19\,$\pm$\,0.03&  1238.4\,$\pm$\,36.40&     3.64\,$\pm$\,0.10\\
           Algol&  LETG&  262.49\,$\pm$\,22.60&  120.90\,$\pm$\,18.00&    0.460\,$\pm$\,0.07&    28.54\,$\pm$\,0.05&  2883.0\,$\pm$\,57.81&     5.11\,$\pm$\,0.17\\
 &   MEG&  71.410\,$\pm$\,9.874&  31.580\,$\pm$\,7.295&    0.442\,$\pm$\,0.11&    28.67\,$\pm$\,0.08&  627.34\,$\pm$\,26.08&     4.32\,$\pm$\,0.23\\
         AR\,Lac&  RGS1&  184.05\,$\pm$\,20.00&  110.15\,$\pm$\,16.82&    0.598\,$\pm$\,0.11&    28.66\,$\pm$\,0.07&  1523.8\,$\pm$\,45.25&     5.14\,$\pm$\,0.23\\
 &   MEG&  16.430\,$\pm$\,4.716&  3.8000\,$\pm$\,2.859&    0.231\,$\pm$\,0.18&    28.37\,$\pm$\,0.28&  247.82\,$\pm$\,16.18&     5.40\,$\pm$\,0.76\\
         AT\,Mic&  RGS1&  501.50\,$\pm$\,26.17&  238.57\,$\pm$\,18.83&    0.475\,$\pm$\,0.04&    27.90\,$\pm$\,0.03&  1603.7\,$\pm$\,43.38&     3.53\,$\pm$\,0.07\\
         AU\,Mic&  RGS1&  944.27\,$\pm$\,36.41&  688.37\,$\pm$\,31.28&    0.729\,$\pm$\,0.04&    27.88\,$\pm$\,0.02&  3224.4\,$\pm$\,61.78&     3.62\,$\pm$\,0.05\\
 &   MEG&  53.790\,$\pm$\,7.523&  45.990\,$\pm$\,6.974&    0.855\,$\pm$\,0.17&    27.58\,$\pm$\,0.07&  454.11\,$\pm$\,21.41&     4.26\,$\pm$\,0.22\\
    $\beta$\,Cet&  RGS1&  128.24\,$\pm$\,14.40&  79.437\,$\pm$\,11.74&    0.619\,$\pm$\,0.11&    28.56\,$\pm$\,0.07&  561.27\,$\pm$\,26.51&     4.01\,$\pm$\,0.18\\
 &  LETG&  241.35\,$\pm$\,21.66&  178.69\,$\pm$\,20.17&    0.740\,$\pm$\,0.10&    28.43\,$\pm$\,0.05&  2176.2\,$\pm$\,50.82&     4.68\,$\pm$\,0.20\\
 &   MEG&  47.120\,$\pm$\,7.198&  23.610\,$\pm$\,5.252&    0.501\,$\pm$\,0.13&    28.19\,$\pm$\,0.10&  548.17\,$\pm$\,23.72&     4.87\,$\pm$\,0.34\\
         Capella&  RGS1&  2611.2\,$\pm$\,60.23&  1520.5\,$\pm$\,46.65&    0.582\,$\pm$\,0.02&    28.53\,$\pm$\,0.01&  7272.0\,$\pm$\,94.37&     3.37\,$\pm$\,0.02\\
 &  LETG&  3071.2\,$\pm$\,56.00&  2135.2\,$\pm$\,51.10&    0.695\,$\pm$\,0.02&    28.49\,$\pm$\,0.01&  14677.\,$\pm$\,124.3&     3.67\,$\pm$\,0.02\\
   $\chi^1$\,Ori&  RGS1&  184.68\,$\pm$\,16.02&  123.42\,$\pm$\,13.40&    0.668\,$\pm$\,0.09&    27.30\,$\pm$\,0.05&  490.75\,$\pm$\,24.35&     3.33\,$\pm$\,0.09\\
         EK\,Dra&  RGS1&  86.306\,$\pm$\,11.43&  59.680\,$\pm$\,9.822&    0.691\,$\pm$\,0.14&    28.16\,$\pm$\,0.07&  409.05\,$\pm$\,22.60&     4.11\,$\pm$\,0.21\\
 &  LETG&  53.253\,$\pm$\,9.516&  31.025\,$\pm$\,7.533&    0.582\,$\pm$\,0.17&    28.12\,$\pm$\,0.11&  324.56\,$\pm$\,19.49&     4.04\,$\pm$\,0.29\\
 $\epsilon$\,Eri&  RGS1&  217.64\,$\pm$\,16.47&  160.21\,$\pm$\,14.34&    0.736\,$\pm$\,0.08&    26.90\,$\pm$\,0.04&  482.75\,$\pm$\,23.73&     3.17\,$\pm$\,0.09\\
 &  LETG&  697.02\,$\pm$\,27.70&  453.99\,$\pm$\,22.78&    0.651\,$\pm$\,0.04&    26.95\,$\pm$\,0.02&  2025.8\,$\pm$\,46.07&     3.13\,$\pm$\,0.05\\
         EQ\,Peg&  RGS1&  273.77\,$\pm$\,19.17&  142.72\,$\pm$\,13.99&    0.521\,$\pm$\,0.06&    27.45\,$\pm$\,0.04&  717.49\,$\pm$\,28.77&     3.31\,$\pm$\,0.07\\
         EV\,Lac&  RGS1&  471.98\,$\pm$\,25.54&  291.00\,$\pm$\,20.31&    0.616\,$\pm$\,0.05&    27.23\,$\pm$\,0.03&  1526.2\,$\pm$\,42.31&     3.55\,$\pm$\,0.07\\
 &   MEG&  119.25\,$\pm$\,11.12&  54.920\,$\pm$\,7.585&    0.460\,$\pm$\,0.07&    27.04\,$\pm$\,0.05&  694.39\,$\pm$\,26.61&     3.71\,$\pm$\,0.15\\
        HR\,1099&  RGS1&  357.12\,$\pm$\,26.86&  254.27\,$\pm$\,23.89&    0.712\,$\pm$\,0.08&    28.73\,$\pm$\,0.04&  2164.1\,$\pm$\,53.29&     4.49\,$\pm$\,0.14\\
 &  LETG&  470.60\,$\pm$\,27.82&  254.79\,$\pm$\,22.49&    0.541\,$\pm$\,0.05&    28.71\,$\pm$\,0.03&  5584.1\,$\pm$\,78.60&     5.25\,$\pm$\,0.12\\
 &   MEG&  207.77\,$\pm$\,16.12&  93.870\,$\pm$\,11.80&    0.451\,$\pm$\,0.06&    28.78\,$\pm$\,0.05&  2867.9\,$\pm$\,54.95&     5.21\,$\pm$\,0.15\\
   $\kappa$\,Cet&  RGS1&  236.88\,$\pm$\,18.54&  153.99\,$\pm$\,15.00&    0.650\,$\pm$\,0.08&    27.34\,$\pm$\,0.04&  632.93\,$\pm$\,27.79&     3.33\,$\pm$\,0.09\\
  $\lambda$\,And&  RGS1&  264.86\,$\pm$\,22.73&  131.08\,$\pm$\,17.88&    0.494\,$\pm$\,0.08&    28.34\,$\pm$\,0.06&  1959.8\,$\pm$\,50.39&     4.92\,$\pm$\,0.20\\
 &   MEG&  74.030\,$\pm$\,9.337&  33.550\,$\pm$\,6.642&    0.453\,$\pm$\,0.10&    28.29\,$\pm$\,0.08&  919.35\,$\pm$\,30.80&     5.02\,$\pm$\,0.28\\
    $\pi^1$\,UMa&  RGS1&  134.95\,$\pm$\,13.36&  94.810\,$\pm$\,11.76&    0.702\,$\pm$\,0.11&    27.39\,$\pm$\,0.06&  383.45\,$\pm$\,21.96&     3.40\,$\pm$\,0.12\\
         Procyon&  LETG&  731.60\,$\pm$\,28.70&  652.40\,$\pm$\,27.40&    0.891\,$\pm$\,0.05&    26.99\,$\pm$\,0.02&  673.20\,$\pm$\,27.60&     2.22\,$\pm$\,0.03\\
   $\sigma$\,CrB&  RGS1&  311.61\,$\pm$\,24.22&  173.44\,$\pm$\,20.53&    0.556\,$\pm$\,0.07&    28.51\,$\pm$\,0.05&  2016.1\,$\pm$\,51.28&     4.62\,$\pm$\,0.16\\
         UX\,Ari&  RGS1&  245.26\,$\pm$\,23.32&  117.54\,$\pm$\,18.79&    0.479\,$\pm$\,0.08&    28.89\,$\pm$\,0.06&  2173.7\,$\pm$\,52.95&     5.27\,$\pm$\,0.20\\
 &  LETG&  300.35\,$\pm$\,22.37&  227.42\,$\pm$\,20.40&    0.757\,$\pm$\,0.08&    28.96\,$\pm$\,0.04&  3321.1\,$\pm$\,61.27&     5.12\,$\pm$\,0.14\\
 &   MEG&  52.410\,$\pm$\,7.733&  20.910\,$\pm$\,5.421&    0.399\,$\pm$\,0.11&    28.92\,$\pm$\,0.10&  524.68\,$\pm$\,23.26&     4.55\,$\pm$\,0.30\\
         VY\,Ari&  RGS1&  205.56\,$\pm$\,19.97&  144.81\,$\pm$\,17.34&    0.704\,$\pm$\,0.10&    28.75\,$\pm$\,0.06&  1314.6\,$\pm$\,40.65&     4.59\,$\pm$\,0.20\\
         YY\,Gem&  LETG&  193.31\,$\pm$\,15.47&  115.75\,$\pm$\,12.44&    0.598\,$\pm$\,0.08&    28.03\,$\pm$\,0.04&  971.47\,$\pm$\,32.27&     3.74\,$\pm$\,0.13\\
         YZ\,CMi&  RGS1&  287.29\,$\pm$\,19.46&  169.54\,$\pm$\,15.80&    0.590\,$\pm$\,0.06&    27.21\,$\pm$\,0.04&  807.37\,$\pm$\,30.60&     3.38\,$\pm$\,0.08\\
\hline
\end{tabular}
\\
$^a$in erg/s\\
$^b$From O\,{\sc viii}/O\,{\sc vii} (Ly/r) ratio with APEC
}
\renewcommand{\arraystretch}{1}
\end{flushleft}
\end{table*}

%% file: tab4.tex
\begin{table}[!ht]
\begin{flushleft}
\renewcommand{\arraystretch}{1.2}
\caption{\label{fr_ne}Measured line counts for Ne\,{\sc ix} resonance (r) and forbidden (f) lines and Ne\,{\sc ix} luminosities L$_{\rm Ne\,\mathsc{ix}}$ from summed (r+i+f) line fluxes.}
{\scriptsize
\begin{tabular}{p{.8cm}p{.7cm}ccc}
 star & Instr.& r & f &log(L$_{\rm Ne\,\mathsc{ix}}$)$^a$\\
         47\,Cas&  RGS2&  452.00\,$\pm$\,38.77&  278.85\,$\pm$\,28.22&    28.64\,$\pm$\,0.04\\
         AB\,Dor&  RGS2&  1614.6\,$\pm$\,57.63&  1030.0\,$\pm$\,50.39&    28.45\,$\pm$\,0.01\\
 &   MEG&  524.16\,$\pm$\,26.51&  290.84\,$\pm$\,18.73&    28.34\,$\pm$\,0.02\\
 &   HEG&  110.99\,$\pm$\,10.98&  77.682\,$\pm$\,9.189&    28.33\,$\pm$\,0.05\\
         AD\,Leo&  RGS2&  553.77\,$\pm$\,37.81&  400.75\,$\pm$\,29.26&    27.22\,$\pm$\,0.03\\
 &  LETG&  276.32\,$\pm$\,19.20&  165.42\,$\pm$\,16.04&    27.16\,$\pm$\,0.04\\
           Algol&  LETG&  631.23\,$\pm$\,36.62&  347.91\,$\pm$\,31.94&    28.83\,$\pm$\,0.04\\
 &   MEG&  283.27\,$\pm$\,23.00&  194.68\,$\pm$\,18.40&    28.65\,$\pm$\,0.04\\
 &   HEG&  79.109\,$\pm$\,9.973&  57.690\,$\pm$\,8.389&    28.79\,$\pm$\,0.06\\
         AR\,Lac&  RGS2&  488.50\,$\pm$\,42.06&  364.31\,$\pm$\,33.39&    29.12\,$\pm$\,0.04\\
 &   MEG&  152.66\,$\pm$\,14.84&  136.82\,$\pm$\,13.30&    29.00\,$\pm$\,0.05\\
 &   HEG&  39.833\,$\pm$\,6.650&  15.984\,$\pm$\,4.306&    28.93\,$\pm$\,0.10\\
         AT\,Mic&  RGS2&  632.39\,$\pm$\,34.03&  387.00\,$\pm$\,28.22&    28.03\,$\pm$\,0.02\\
         AU\,Mic&  RGS2&  1156.9\,$\pm$\,45.71&  812.28\,$\pm$\,41.49&    27.97\,$\pm$\,0.02\\
 &   MEG&  319.98\,$\pm$\,19.79&  212.47\,$\pm$\,15.78&    27.74\,$\pm$\,0.03\\
 &   HEG&  87.094\,$\pm$\,9.526&  52.332\,$\pm$\,7.293&    27.80\,$\pm$\,0.06\\
    $\beta$\,Cet&  LETG&  679.75\,$\pm$\,41.78&  465.82\,$\pm$\,35.45&    28.78\,$\pm$\,0.03\\
 &   MEG&  363.20\,$\pm$\,23.78&  211.92\,$\pm$\,17.14&    28.58\,$\pm$\,0.03\\
 &   HEG&  82.103\,$\pm$\,9.361&  64.720\,$\pm$\,8.301&    28.61\,$\pm$\,0.06\\
         Capella&  LETG&  3960.8\,$\pm$\,79.78&  2312.8\,$\pm$\,66.94&    28.58\,$\pm$\,0.01\\
 &   MEG&  2512.9\,$\pm$\,60.13&  1373.2\,$\pm$\,42.87&    28.43\,$\pm$\,0.01\\
 &   HEG&  616.43\,$\pm$\,25.70&  373.48\,$\pm$\,20.09&    28.46\,$\pm$\,0.02\\
         EK\,Dra&  RGS2&  172.87\,$\pm$\,20.22&  139.63\,$\pm$\,18.39&    28.47\,$\pm$\,0.05\\
 $\epsilon$\,Eri&  RGS2&  213.28\,$\pm$\,19.03&  125.74\,$\pm$\,16.75&    26.87\,$\pm$\,0.04\\
 &  LETG&  564.02\,$\pm$\,27.20&  347.98\,$\pm$\,22.35&    26.82\,$\pm$\,0.02\\
         EQ\,Peg&  RGS2&  209.59\,$\pm$\,23.72&  154.66\,$\pm$\,18.08&    27.41\,$\pm$\,0.05\\
         EV\,Lac&  RGS2&  376.34\,$\pm$\,31.23&  253.93\,$\pm$\,24.17&    27.14\,$\pm$\,0.04\\
 &   MEG&  454.53\,$\pm$\,23.74&  222.32\,$\pm$\,16.34&    27.02\,$\pm$\,0.03\\
 &   HEG&  115.34\,$\pm$\,11.01&  68.823\,$\pm$\,8.547&    27.08\,$\pm$\,0.05\\
        HR\,1099&  RGS2&  762.77\,$\pm$\,44.32&  409.12\,$\pm$\,36.44&    29.03\,$\pm$\,0.03\\
 &  LETG&  1231.7\,$\pm$\,45.61&  727.60\,$\pm$\,39.44&    29.09\,$\pm$\,0.02\\
 &   MEG&  1690.5\,$\pm$\,50.43&  834.17\,$\pm$\,34.81&    29.13\,$\pm$\,0.01\\
 &   HEG&  477.47\,$\pm$\,22.85&  228.75\,$\pm$\,16.25&    29.21\,$\pm$\,0.02\\
   $\kappa$\,Cet&  RGS2&  251.60\,$\pm$\,23.27&  193.82\,$\pm$\,21.53&    27.40\,$\pm$\,0.04\\
  $\lambda$\,And&  RGS2&  532.86\,$\pm$\,39.36&  335.00\,$\pm$\,33.78&    28.71\,$\pm$\,0.03\\
 &   MEG&  506.89\,$\pm$\,26.09&  219.34\,$\pm$\,17.80&    28.55\,$\pm$\,0.03\\
 &   HEG&  54.933\,$\pm$\,7.704&  35.636\,$\pm$\,6.216&    28.28\,$\pm$\,0.08\\
    $\pi^1$\,UMa&  RGS2&  158.38\,$\pm$\,19.24&  105.83\,$\pm$\,15.87&    27.46\,$\pm$\,0.06\\
         Procyon&  LETG&  66.930\,$\pm$\,10.46&  35.030\,$\pm$\,8.468&    25.83\,$\pm$\,0.10\\
   $\sigma$\,CrB&  RGS2&  693.59\,$\pm$\,48.82&  344.81\,$\pm$\,37.72&    28.86\,$\pm$\,0.03\\
         UX\,Ari&  RGS2&  671.37\,$\pm$\,40.05&  456.80\,$\pm$\,33.42&    29.42\,$\pm$\,0.02\\
 &  LETG&  564.02\,$\pm$\,27.20&  347.98\,$\pm$\,22.35&    29.19\,$\pm$\,0.02\\
 &   MEG&  409.21\,$\pm$\,23.32&  237.38\,$\pm$\,17.32&    29.33\,$\pm$\,0.03\\
 &   HEG&  48.597\,$\pm$\,7.493&  18.571\,$\pm$\,4.596&    28.95\,$\pm$\,0.10\\
         VY\,Ari&  RGS2&  481.21\,$\pm$\,33.86&  306.33\,$\pm$\,28.95&    29.11\,$\pm$\,0.03\\
         YY\,Gem&  LETG&  240.53\,$\pm$\,18.54&  137.45\,$\pm$\,15.34&    28.06\,$\pm$\,0.05\\
         YZ\,CMi&  RGS2&  195.54\,$\pm$\,22.57&  101.04\,$\pm$\,15.84&    27.01\,$\pm$\,0.06\\
\hline
\end{tabular}
\\
$^a$in erg/s\\
}
\renewcommand{\arraystretch}{1}
\end{flushleft}
\end{table}